\documentclass[11pt,export]{article}
\usepackage[st,hs,english]{ethasl}

\usepackage[
  backend=biber,
  style=ieee,
  eprint=true,
  arxiv=abs
]{biblatex}
\usepackage[T1]{fontenc}

\usepackage{a4}

\usepackage{fancyhdr}

\usepackage{textcomp}\usepackage{gensymb}

\usepackage[english]{isodate}

\usepackage{multicol}

\usepackage{graphicx}
\usepackage{float}

\usepackage{tocloft}

\usepackage{booktabs}
\usepackage{array}
\usepackage{multirow}

\usepackage{color}
\usepackage{colortbl}
\usepackage[most]{tcolorbox}
\definecolor{black}{rgb}{0,0,0}
\definecolor{white}{rgb}{1,1,1}
\definecolor{darkred}{rgb}{0.5,0,0}
\definecolor{darkgreen}{rgb}{0,0.5,0}
\definecolor{darkblue}{rgb}{0,0,0.5}

\usepackage{amsmath}
\usepackage{amssymb}
\usepackage{cancel}
\usepackage{mathtools}

\usepackage{nicefrac}

\usepackage{upgreek}

\usepackage{isomath}

\usepackage{units}

\usepackage{rotating}

\usepackage{pdfpages}
\includepdfset{pages={-}, frame=true, pagecommand={\thispagestyle{fancy}}}

\usepackage{pgfgantt}

\usepackage[hidelinks]{hyperref}
\hypersetup{
  colorlinks   = true, 
  urlcolor     = blue, 
  linkcolor    = [RGB]{31, 81, 255}, 
  citecolor   = [RGB]{255, 68, 51} 
}

\usepackage{xcolor}

\usepackage{cleveref}

\usepackage{array}
\usepackage{hhline}

\usepackage{multirow}

\usepackage{feynmp-auto}

\usepackage[mode=buildnew]{standalone}

\usepackage{appendix}

\usepackage{physics}

\usepackage{adjustbox}

\usepackage{fancyhdr}

\usepackage{caption}
\usepackage{subcaption}

\usepackage[a4paper, portrait, margin=1in]{geometry}
\usepackage{slashed}
\usepackage{enumitem}

\newcommand{\VP}[1]{\textcolor{BrickRed}{#1}}

\title{\bfseries\boldmath

  Two-loop BRST-restoring finite counter-terms for chiral non-abelian gauge theories
}
\author{
  Achilleas Lazopoulos and Vera Paroutiadou\\[0.5em]
  {\small\itshape Institute for Theoretical Physics, ETH Zurich,}\\
  {\small\itshape 8093 Z\"urich, Switzerland}
}
\date{}
\begin{document}
\maketitle
\begin{abstract}
We study the renormalization of anomaly-free chiral $SU(N)$
gauge theories without scalar fields through two-loop order.
We adopt the Breitenlohner--Maison--'t Hooft--Veltman (BMHV)
scheme for the treatment of $\gamma_5$ in dimensional
regularization. This prescription induces spurious breaking
of BRST symmetry, requiring finite symmetry-restoring
counter-terms in addition to the subtraction of ultraviolet
divergences. We derive these counter-terms for a general,
possibly reducible fermion representation,  thus restoring the
BRST symmetry, through two loops.
\end{abstract}

\tableofcontents

\section{Introduction}
Handling chiral gauge theories within Dimensional Regularization has a long and oscillating history. 
The problem of analytically continuing  the Dirac algebra to $d$-dimensions, in view of $\gamma_5$'s inherent four-dimensional nature was already identified in the original work by 't Hooft and Veltman\cite{tHooft:1971akt,tHooft:1972tcz}. Their approach of keeping $\gamma_5$ strictly four-dimensional was consolidated by Breitenlohner and Maison~\cite{Breitenlohner:1975hg,Breitenlohner:1976te,Breitenlohner:1977hr} into the Breitenlohner-Maison-t Hooft-Veltman (BMHV) scheme, which is algebraically consistent but practically cumbersome. The complexity introduced has motivated attempts to find shortcuts by postulating a fully anticommuting $\gamma_5$ at the expense of sacrificing algebraic operations, like trace cyclicity, as in work by Kreimer et al.~\cite{Kreimer:1989ke,Korner:1991sx,Kreimer:1993bh}, further developed in many variations~\cite{gamma5_facts_life,Chen:2024hlv} which are collectively\footnote{and somewhat unfairly} referred to as "Naive Dimensional Regularization" (NDR). In all these approaches one must employ a (process dependent) physical constraint to a posteriori fix the ambiguous result of an amplitude computation, like, for example, anomaly cancellation, Bose symmetry in case it applies, hermiticity of the effective action, symmetry identities etc. or alternatively, fix the ambiguity of traces by reading point prescriptions that have been verified to give the correct result. A collection of reading point prescriptions that is sufficient for all cases necessary when deriving two loop RGEs in the SMEFT can be found in~\cite{Born:2026xkr}.
This has worked for many specific cases at one loop and, in some cases, well beyond, but, as can be seen in ~\cite{Bednyakov:2015ooa,Davies:2019onf} surprises might appear due to specific trace configurations for the first time at high loop level. 

The alternative prescription by Larin~\cite{Larin:1993tq} preserves the four-dimensional nature of the axial current, but also needs finite renormalizations to re-impose Ward identities. Employing an altogether different regularization scheme, like implicit fixed dimension methods such as CDR, CIReg and FDR (see \cite{Gnendiger:2017pys} for a review), has been shown to suffer from similar problems~\cite{Bruque:2018bmy} that can also be fixed by finite symmetry restoring counter-terms. 

While the BMHV scheme has the merit of being algebraically sound, it introduces an explicit breaking 
of gauge and (in general gauges) BRST invariance which, although of evanescent nature, induces finite, four-dimensional 
effects in amplitudes or Green functions that are UV singular. 
The remedy is the introduction of finite, symmetry restoring counter-terms, that have to be computed and added to the effective action order by order in the perturbative expansion, and whose precise form depends on the particle content of the theory. 

These counter-terms restore the symmetry at the level of the Effective Action, and, consequently, also the Slavnov-Taylor identities that depend on the BRST symmetry. The procedure of computing them is somewhat simplified if one is 
willing to work in the Background Field Gauge (BFG), which preserves gauge invariance in the Effective Action. Thereby what one seeks to restore are the simpler Ward identities. This was done at the one loop level for chiral theories~\cite{Cornella:2022hkc}. Recently Olgoso and Vecchi proposed the usage of spurion fields, 
as is done in chiral perturbation theory, to compute one loop symmetry restoring counter-terms in the BFG~\cite{OlgosoRuiz:2024dzq}, which was later automated and extended to the entire SMEFT~\cite{Fuentes-Martin:2025meq}. 
Mapping between NDR and BMHV schemes and identifying cases when the difference vanishes in matching calculations
is the topic of~\cite{DiNoi:2025uan}.

For covariant gauges, one loop symmetry restoring counter-terms for chiral gauge theories have been computed in~\cite{Martin:1999cc,Sanchez-Ruiz:2002pcf}. 

Symmetry restoring counter-terms for BMHV computations in specific processes have been found on a case-per-case basis~\cite{Heller:2020owb}. In connection to the renormalization program of effective field theories, BMHV counter-terms relevant to the LEFT at one-(\cite{Naterop:2023dek}) and two-loop level~\cite{Naterop:2023dek,Naterop:2024cfx,Naterop:2025cwg,Naterop:2025lzc} have been derived. In the context of the SMEFT, symmetry restoring counter-terms for dim-6  operators have been reported at one loop~\cite{Solera:2026xnz} and a comparison between different schemes in the context of four-fermion operators in gluon fusion has been discussed, earlier, in~\cite{DiNoi:2023ygk}. 

One loop counter-terms for chiral abelian Yang-Mills were published in~\cite{Belusca-Maito:2020jiz}. 
Two loop RGEs and symmetry restoring counter-terms for chiral abelian Yang-Mills were 
published in~\cite{Belusca-Maito:2022wem,Belusca-Maito:2021lnk}(2021), see also ~\cite{Belusca-Maito:2023wah} for a review of the approach, and even 3-loop RGEs for chiral abelian theories were computed in ~\cite{Stockinger:2023ndm} and four-loops in ~\cite{vonManteuffel:2025swv}, see also~\cite{Weisswange:2025tab}. In ~\cite{Ebert:2024xpy}the authors explore how symmetry restoring counter-terms
change when varying the choices of how to continue the Lagrangian from 4 to d-dimensions within BMHV. Very recently symmetry restoring counter-terms were computed also for massive gauge theories at one loop~\cite{Quadri:2026zwv}. 

Symmetry restoring counter-terms within the BMHV scheme, for an $SU(2)$ chiral theory, at the one-loop level were presented in~\cite{Belusca-Maito:2020ala} and at the two loop level in~\cite{Kuhler:2025znv}. 
In this work we follow the approach of St\"ockinger et al. aiming to extend the results of ~\cite{Kuhler:2025znv}, to which we compare repeatedly. Thus, we compute the BRST violating effects to the effective action, using insertions of local operators, and then computing the symmetry restoring counter-terms whose BRST variation 
cancels those contributions. 
We use a variant of the R-operation with an IR regulator mass to compute UV singular terms and the corresponding BRST violating finite terms diagram by diagram, at the one- and two- loop level, for a chiral $SU(N)$ gauge theory in which only the right-handed fermion components participate in the gauge interaction. This implies that some of the details of the algorithmic approach of~\cite{Kuhler:2025znv} have to be modified in our case - in particular the one-loop insertion operators used at the two-loop order remain evanescent. 
We keep the representation of the fermions arbitrary, allowing for reducible representations, but we assume a gauge anomaly cancellation condition, so that the theory is well-defined~\cite{Witten:1982fp}. 
We restrict the Lagrangian to operators up to mass dimension four. 
The calculation is performed in the covariant gauge. We provide results for arbitrary $\xi$ at the one-loop level,
but restrict ourselves to $\xi=1$ at  two-loops. All results are presented in $\overline{\rm MS}$-scheme. 

The paper is organized as follows. We first define the theory, introduce the BMHV conventions, and identify the origin of the tree-level evanescent breaking. We then summarize the BRST restoration machinery and the renormalization procedure used throughout the calculation. The one-loop analysis is presented next, including the singular counter-terms, the Slavnov-Taylor breaking, the finite restoring counter-terms, and the associated consistency check. Finally, we extend the construction to two loops, where the full set of ultraviolet and finite symmetry-restoring contributions is derived.

\section{Setting up the theory}
\label{ch:Section1}
We start by introducing the massless tree-level Lagrangian of a chiral
$SU(N)$ gauge theory containing fermionic matter fields and no scalars,
while establishing the conventions used throughout this work.
We consider only right-handed fermions $\psi_{Ri}$, which couple to
the gauge fields $G_\mu^a$ and form a multiplet transforming in a
generic, possibly reducible, representation
$\mathcal R=\bigoplus_{m=1}^{M}R_m$ of $SU(N)$.
Each block $R_m$ is an irreducible representation, and repeated
copies are included as separate blocks. The generators are
\begin{equation}
T_{\mathcal R}^a
=
\operatorname{diag}
\left(T_{R_1}^a,\ldots,T_{R_M}^a\right),
\qquad
[T_{\mathcal R}^a,T_{\mathcal R}^b]
=
i f^{abc}T_{\mathcal R}^c,
\end{equation}
where $f^{abc}$ are the structure constants of the gauge group.
The indices $i,j$ run over all components of the fermion multiplet,
$i,j=1,\ldots,\dim\mathcal R$, with
$\dim\mathcal R=\sum_m\dim R_m$.

We assume cancellation of the perturbative local gauge anomaly, which is the condition relevant for the local Slavnov-Taylor analysis performed here. 
\begin{equation}
    \operatorname{Tr}_R\!\left(T_R^a\{T_R^b,T_R^c\}\right)=0.
\end{equation}
Individual irreducible blocks do not need to satisfy this condition
separately. Possible global gauge anomalies, in particular the $SU(2)$ anomaly, are outside the scope of the present perturbative analysis.

For the purposes of dimensional regularization, each right-handed
Weyl component is embedded into a Dirac spinor,
\begin{equation}
\psi_i=\psi_{Li}+\psi_{Ri},
\qquad
\psi_{Li}=\mathbb P_L\psi_i,
\qquad
\psi_{Ri}=\mathbb P_R\psi_i.
\end{equation}
Here the chiral projectors are defined by
\[
\mathbb P_R=\frac{1+\gamma_5}{2},
\qquad
\mathbb P_L=\frac{1-\gamma_5}{2}.
\]
The left-handed components $\psi_{Li}$ are fictitious gauge singlets
that do not participate in the physical gauge interactions.

To define the chiral theory consistently in $d$--dimensions, we adopt the Breitenlohner--Maison--'t Hooft--Veltman (BMHV) scheme, with
\begin{equation}
d=4-2\epsilon.
\end{equation}
Within this framework, the $d$--dimensional spacetime is decomposed into a four-dimensional physical subspace and a $d-4$--dimensional evanescent complement. Objects associated with these subspaces are denoted by barred and hatted quantities, respectively. In particular,
\begin{equation}
g_{\mu\nu}
=
\overline{g}_{\mu\nu}
+
\widehat{g}_{\mu\nu},
\qquad
\overline{g}^{\mu}_{\mu}=4,
\qquad
\widehat{g}^{\mu}_{\mu}=d-4=-2\epsilon,
\qquad
\overline{g}_{\mu\rho}\widehat{g}^{\rho\nu}=0.
\end{equation}
Accordingly, the Dirac matrices decompose as
\begin{equation}
\gamma^\mu
=
\overline{\gamma}^\mu
+
\widehat{\gamma}^\mu.
\end{equation}
The four-dimensional anticommutation relation between $\gamma_5$ and the Dirac matrices is retained only in the physical subspace. The defining BMHV relations are
\begin{equation}
\label{eqn:anticom}
\left\{\gamma_5,\overline{\gamma}^{\mu}\right\}=0,
\qquad
\left[\gamma_5,\widehat{\gamma}^{\mu}\right]=0,
\qquad
\left\{\gamma_5,\gamma^\mu\right\}
=
2\widehat{\gamma}^{\mu}\gamma_5.
\end{equation}
With these conventions, we choose the $d$--dimensional continuation\footnote{The continuation of the fermion--gauge interaction to $d$ dimensions
is not unique. We choose an interaction containing only barred
Lorentz components. An alternative continuation may include
an evanescent interaction proportional to an arbitrary parameter
$\alpha$. Its BRST variation modifies the tree-level BRST breaking and can introduce additional insertion vertices,
depending on the chosen interaction.} of the tree-level Lagrangian as
\begin{equation}
\mathcal L
=
\mathcal L_{\rm YM}
+\mathcal L_{\rm F}
+\mathcal L_{\rm GF}
+\mathcal L_{\rm ghost}
+\mathcal L_{\rm ext}.
\end{equation}
with
\begin{eqnarray}
\mathcal L_{\rm YM}
&=&
-\frac14 G_{\mu\nu}^a G^{a,\mu\nu}
\\
\mathcal L_{\rm F} &=&
i\overline{\psi}_i\overline{\slashed{\partial}}\psi_i 
+ \VP{
 i\overline{\psi}
\widehat{\slashed{\partial}}\psi
}
+ g_s T_{\mathcal R,ij}^a\,
\overline\psi_i\,
\overline{\slashed G}^{\,a}\mathbb P_R\psi_j 
\\
\mathcal L_{\rm GF} & = & 
\frac{\xi}{2}(B^a)^2
+B^a\partial^\mu G_\mu^a
\\
\mathcal{L_{\rm ghost}} & = &
\overline c^{\,a}
\left(-\partial^\mu D_\mu^{ac}\right)c^c
\\
\mathcal L_{\rm ext} &=& 
\rho_a^{\mu}s_dG_\mu^a
+
\zeta^a s_dc^a
+
\chi^a s_d\overline{c}^{a}
+
\overline{R}^{i}s_d\psi_i
+
R^i s_d\overline{\psi}_i.
\end{eqnarray}

Here,
\begin{equation}
G_{\mu\nu}^a
=
\partial_\mu G_\nu^a
-\partial_\nu G_\mu^a
+g_s f^{abc}G_\mu^bG_\nu^c,
\end{equation}
and the covariant derivatives in the fermionic and adjoint
representations are defined by
\begin{equation}
D_{\mu,ij}
=
\partial_\mu\delta_{ij}
-i g_s G_\mu^a T_{\mathcal R,ij}^a,
\qquad
D_\mu^{ac}
=
\partial_\mu\delta^{ac}
+g_s f^{abc}G_\mu^b.
\end{equation}

We identify $c^a$ and $\overline c^a$ as the ghost and antighost fields and $B^a$ as the usual auxiliary Nakanishi--Lautrup field.

We have introduced external sources $\rho,\zeta,\chi,\overline{R},R$ in order to formulate the Slavnov-Taylor identity and to control the renormalization of the composite operators generated by nonlinear BRST transformations. The quantum numbers and canonical dimensions of the external sources
are chosen such that every term in $\mathcal L_{\rm ext}$ has ghost
number zero and the appropriate mass dimension. The explicit form of $s_d G_\mu^a,\ldots,s_d\overline{\psi}_i$ is given in section~\ref{section:BRST_general}.

Additionally, we have chosen to work in a general covariant gauge parametrized by the gauge parameter $\xi$.

Note that the fermion--gauge interaction is chosen, for convenience, to lie entirely in the physical 
four-dimensional subspace (as specified by the bar in $g_s T_{\mathcal R,ij}^a\,
\overline\psi_i\,
\overline{\slashed G}^{\,a}\mathbb P_R\psi_j$). 
The gauge field itself remains $d$--dimensional and the projection onto the four-dimensional subspace concerns only the Lorentz structure entering the chiral fermion--gauge vertex:
\begin{equation}
g_s \,
\overline\psi_i\,
\overline{\slashed G}^{\,a}T_{\mathcal R,ij}^a\mathbb P_R\psi_j =   
g_s \,
\overline\psi_i\,
\overline{\gamma}^\mu G_{\mu}^{\,a}T_{\mathcal R,ij}^a\mathbb P_R\psi_j
\end{equation}

The $d$-dimensional continuation of the fermionic kinetic term, on the other hand, necessarily contains an evanescent contribution, since we want to keep working with d-dimensional loop integrals. Using
the decomposition of gamma matrices characterizing the BMHV scheme, one obtains
\begin{equation}
\label{eqn:fermion-kinetic-decomposition}
\begin{aligned}
\mathcal{L}_{\rm ferm,kin}
&=
i\overline{\psi}_i
\left(
\overline{\gamma}^{\mu}
+
\widehat{\gamma}^{\mu}
\right)
\partial_\mu\psi_i
\\[0.5em]
&=
i\overline{\psi}_i
\overline{\slashed{\partial}}\psi_i
+
i\overline{\psi}_i
\widehat{\slashed{\partial}}\psi_i
\\[0.5em]
&=
i\overline{\psi}_i
\overline{\slashed{\partial}}
\left(\mathbb{P}_L+\mathbb{P}_R\right)\psi_i
+
i\overline{\psi}_i
\widehat{\slashed{\partial}}
\left(\mathbb{P}_L+\mathbb{P}_R\right)\psi_i
\\[0.5em]
&=
i\overline{\psi}_{Li}
\overline{\slashed{\partial}}
\psi_{Li}
+
i\overline{\psi}_{Ri}
\overline{\slashed{\partial}}
\psi_{Ri}
+
\VP{i\overline{\psi}_{Ri}
\widehat{\slashed{\partial}}
\psi_{Li}}
+
\VP{i\overline{\psi}_{Li}
\widehat{\slashed{\partial}}
\psi_{Ri}}.
\end{aligned}
\end{equation}
The first two terms preserve the separation between the left- and right-handed sectors. In contrast, the last two terms mix the gauge-singlet left-handed components with the gauge-transforming right-handed components. They therefore constitute an evanescent, tree-level breaking of gauge invariance and, correspondingly, of BRST symmetry in the $d$-dimensional regularized theory. The analysis that follows is devoted exclusively to the consequences of this regularization-induced BRST breaking and to its restoration at the level of renormalized Green functions.

\section{BRST restoration machinery}
\label{section:BRST_general}
The 4-dimensional Lagrangian is invariant under a global BRST transformation with a Grassmann parameter $\theta$. The 4-dimensional BRST operator, $s_4$ is defined by
\begin{equation}
\delta_\theta \phi_i = \theta\; s_4(\phi_i)
\end{equation}
We can continue the transformation to d-dimensions and define the 
 $d$-dimensional BRST operator $s_d$, which, under our conventions
acts on the individual fields as 
\begin{align}
s_d G_\mu^a
&=
\partial_\mu c^a
+g_s f^{abc}G_\mu^b c^c,
\\[0.5em]
s_d\psi_i
&=
i g_s T_{\mathcal R,ij}^a c^a\psi_{Rj},
\\[0.5em]
s_d\overline\psi_i
&=
i g_s T_{\mathcal R,ji}^a\,
\overline\psi_{Rj}c^a,
\\[0.5em]
s_dc^a
&=
-\frac{g_s}{2}f^{abc}c^b c^c,
\\[0.5em]
s_d\overline c^{\,a}
&=
B^a,
\\[0.5em]
s_dB^a
&=
0.
\end{align}

The external BRST sources are taken to be BRST invariant,
\begin{equation}
s_d X_i=0,
\qquad
X_i\in
\left\{
\rho_a^\mu,\zeta^a,\chi^a,\overline R^i,R^i
\right\}.
\end{equation}

Since the BRST transformations are nonlinear, quantum BRST
invariance is expressed by the Slavnov--Taylor identity for the
effective action. The BMHV regularization breaks this identity,
which must be restored by suitable counterterms.

Acting on a
Grassmann-even functional $\mathcal F$, the Slavnov-Taylor functional is defined as
\begin{equation}
\label{eq:ST_operator}
\begin{aligned}
\mathcal S_d(\mathcal F)
= \frac{\delta\mathcal F}{\delta X_i}
\circ
\frac{\delta\mathcal F}{\delta\phi_i} =
\int d^d x\,\Bigg(
&
\frac{\delta\mathcal F}{\delta\rho_a^\mu}
\frac{\delta\mathcal F}{\delta G_\mu^a}
+
\frac{\delta\mathcal F}{\delta\zeta_a}
\frac{\delta\mathcal F}{\delta c^a}
+
\frac{\delta\mathcal F}{\delta\overline R^i}
\frac{\delta\mathcal F}{\delta\psi_i}
+
\frac{\delta\mathcal F}{\delta R^i}
\frac{\delta\mathcal F}{\delta\overline\psi_i}
+
B^a\frac{\delta\mathcal F}{\delta\overline c^a}
\Bigg).
\end{aligned}
\end{equation}
The subscript $d$ denotes the spacetime dimension.
We use left functional derivatives throughout, defined by
\begin{equation}
\delta\mathcal F
=
\int d^dx\,\delta\Phi(x)\,
\frac{\delta_L\mathcal F}{\delta\Phi(x)},
\end{equation}
with summation over fields and sources understood.
The factors in Eq.~\eqref{eq:ST_operator} are ordered as displayed.

Note that the source $\chi^a$,
coupled to the linear transformation $s_d\overline c^{\,a}=B^a$,
may be retained for completeness, although only sources coupled to
nonlinear BRST transformations are required for the standard
algebraic-renormalization construction.

At tree level, in 4-dimensions, the statement that the action is BRST invariant is equivalent 
to the vanishing of the Slavnov--Taylor functional evaluated
on the classical 4-dimensional action: 
\begin{equation}
\mathcal S_4(S_{0;d=4}) = s_d S_{0,d=4} = 0
\end{equation}
where $S_{0;d=4} = \int d^4x \mathcal{L}|_{d=4}$ is the action restricted to 4-dimensions, 
i.e. with all evanescent structures set to zero. 

For $\mathcal F_{\rm sp}=S_0+F$, where $F$ is independent
of the external BRST sources, the source couplings retain
their classical form. Consequently,
\begin{equation}
\mathcal S_d(\mathcal F_{\rm sp})
=
s_d\mathcal F_{\rm sp}.
\end{equation}
In $d$-dimensions, the action of $\mathcal{S}_d$ on $S_0 = \int d^dx \mathcal L$ with $\mathcal L$ defined above, reproduces its BRST variation:
\begin{equation}
\begin{aligned}
\label{eq:ST_tree}
\widehat\Delta &\equiv  
\,\mathcal S_d(S_0)
=
s_dS_0 
= 
s_d 
\int d^dx \, i\left(
\VP{
\overline{\psi}_{i}
\widehat{\slashed{\partial}}
\mathbb{P}_{R}\psi_{i}
+
\overline{\psi}_{i}\mathbb{P}_{L}
\widehat{\slashed{\partial}}
\psi_{i}
}
\right) \\
&=
\int d^{d}x\,
g_s T^a_{\mathcal R,ij} c^a
\left\{
\overline{\psi}_{i}
\overleftarrow{\widehat{\slashed{\partial}}}
\mathbb{P}_{R}\psi_{j}
+
\overline{\psi}_{i}\mathbb{P}_{L}
\overrightarrow{\widehat{\slashed{\partial}}}
\psi_{j}
\right\},
\end{aligned}
\end{equation}

where $\widehat\Delta$ denotes the evanescent breaking
induced by the BMHV continuation of the fermionic sector.

\par
Beyond tree level, the nonlinear structure of the
Slavnov--Taylor functional couples different perturbative
orders. Consider, for instance,
\begin{equation}
\mathcal F=S_0+\mathcal F_1,
\end{equation}
where $\mathcal F_1$ denotes a contribution of one-loop
order. Expanding to linear order in $\mathcal F_1$ and
using the classical source couplings, we obtain
\begin{equation}
\mathcal S_d(S_0+\mathcal F_1)
=
s_dS_0+s_d\mathcal F_1
+
\int d^d x\,\Bigg(
\frac{\delta\mathcal F_1}{\delta\rho_a^\mu}
\frac{\delta S_0}{\delta G_\mu^a}
+
\frac{\delta\mathcal F_1}{\delta\zeta_a}
\frac{\delta S_0}{\delta c^a}
+
\frac{\delta\mathcal F_1}{\delta\overline R^i}
\frac{\delta S_0}{\delta\psi_i}
+
\frac{\delta\mathcal F_1}{\delta R^i}
\frac{\delta S_0}{\delta\overline\psi_i}
\Bigg)
+\mathcal O(\mathcal F_1^2).
\end{equation}
For source-dependent functionals, the linear contribution
therefore contains additional terms beyond the field
BRST variation $s_d\mathcal F_1$. At higher perturbative
orders, the quadratic part of the Slavnov--Taylor functional
also contributes through products of lower-order terms.
\par
Under the  gauge-anomaly cancellation assumption
the condition for BRST restoration is
\begin{equation}
\label{eq:ST_restoration}
\lim_{d\rightarrow4}
\mathcal S_d\left(\Gamma_{\rm ren}\right)
=0,
\end{equation}
imposed order by order in perturbation theory.
Here the limit includes the projection onto the physical
four-dimensional subspace, while $\Gamma_{\rm ren}$ denotes
the renormalized quantum effective action. 
The central objects of this work are the symmetry restoring counter-terms that 
we need to add to the bare effective action in order to satisfy the condition of eq.~\ref{eq:ST_restoration}
at each order in perturbation theory.

The evanescent breaking of BRST at tree-level induces further breaking of BRST
at the loop level, order by order in perturbation theory, 
which is not exclusively evanescent any more, 
but also 4-dimensional. As evanescent contributions in loop integrals lead 
to  $\epsilon$-terms in numerators that are multiplied by UV poles, new, 
finite and 4-dimensional contributions appear in the effective action, that need 
to be canceled by explicit, finite counter-terms that should be added to the
effective action, to restore eq.~\ref{eq:ST_restoration}. As this effect follows
the renormalization procedure, the difficulty in computing finite restoring counter-terms 
depends on the renormalization procedure followed, to which we turn our attention next.

\section{UV renormalization}
\label{sec:renormalization-bmhv}
Beyond one loop, BRST restoration, like renormalization itself,
is complicated by the presence of nested singularities
from divergent subgraphs. In this section, we outline the renormalization procedure employed throughout this work. 

We compute the UV poles (and the corresponding BRST variation terms) that every individual diagram contributes to the renormalized effective action, in a BPHZ manner, 
carefully taking into account
sub-divergences that are canceled by counter-terms one loop lower. Specifically, we employ a variant of the $R$-operation which uses a mass regulator, $M$, to shield us from IR 
singularities. 
Using a fictitious mass to regulate IR singularities and  expansion identities to isolate the UV contributions of diagrams has been used, as alternative to $R^*$~\cite{Chetyrkin:1982nn,Chetyrkin:1984xa}, in various context before, see e.g.~\cite{Misiak_1995}, the three-loop beta function computation by Chetyrkin et al. in ~\cite{Chetyrkin:1997fm} but also~\cite{Hollik:2002mv} and~\cite{Stockinger:2023ndm,vonManteuffel:2025swv}.
The particular combination of $R$-operation with a mass regulator is closely related to the established auxiliary-mass infrared-rearrangement literature, but as far as we know, employed explicitly for the 
first time in~\cite{Capatti:2022tit}  while a closely related approach was described in~\cite{Pozzorini:2020hkx,Lang:2020nnl} (see appendix of ~\cite{Lang:2020nnl}). We schematically describe the insertion of the regulator mass in the expression for a diagram $\gamma \to \tilde{\gamma}(M,\lambda)$, as a replacement, in the denominator of every propagator 
\begin{equation}
\frac{1}{(c_i k_i+q_i)^2-m_i^2}
\;\longrightarrow\;
\frac{1}{
(c_i k_i+\lambda q_i)^2
-\lambda^2m_i^2
-M^2+\lambda^2M^2
}.
\end{equation}
together with the rescaling of all external (to the diagram\footnote{If we are computing a sub-diagram, the loop momenta of loops outside the sub-diagram are considered external.}) momenta and all masses, in numerator and denominators. For a two-loop diagram $\gamma$ with proper
divergent one-loop subgraphs $\gamma_i$, the overall pole contribution has the form
\begin{equation}
\label{eq:z_operation}
    \mathcal{Z}(\tilde{\gamma}) - \sum_i\mathcal{Z}\left( \tilde{\gamma} / \tilde{\gamma}_i \ast \mathcal{Z}(\tilde{\gamma}_i) \right)
\end{equation}
where $\tilde{\gamma}/\tilde{\gamma}_i$ is the diagram that remains when $\gamma_i$ is shrunk to a point inside $\gamma$. Note that the actual UV poles of the diagram also contain a term $+\sum_i \tilde{\gamma} / \tilde{\gamma}_i \ast \mathcal{Z}(\tilde{\gamma}_i)$, which is non-local (it's a sum of insertions of  lower loop counter-terms in  higher loop diagrams), but cancels with lower order diagrams with counter-term insertions, following BPHZ combinatorics.  

In the above expression, $\mathcal{Z}(X)$ stands for the operation of Taylor expanding the expression in $\lambda$ up to the order equal to the superficial degree of divergence $\omega$ of $X$, denoted by $\mathcal{T}_\omega$, then performing the loop integral(s), retaining only the pole part, denoted by $\mathcal{K}_\epsilon$, and finally setting $\lambda \to 1$:
\begin{equation}
    \mathcal{Z}(X) = \lim_{\lambda\to 1}\mathcal{K}_\epsilon\left(\mathcal{T}_\omega(X)\right)
\end{equation}

This is, in essence, a hard-region expansion of the expression for $X$, to the order specified by the degree of divergence. Note that, thanks to the presence of the IR regulator mass, $M$, there are no scaleless integrals to consider, and, also, the only integrals necessary after tensor reduction are tadpole integrals with a common mass $M$. However, we also note that there are ample cancellations of $M$ between the numerator and the result of the integral computation, and that these terms must be duly retained. 

Although individual terms of the expression in eq.~\ref{eq:z_operation} depend on $M$, the whole expression does not, and this is a valuable test when implementing the algorithm. We stress that this is not necessarily the case for finite terms in the regular hard region expansion. It is sufficient, however, to isolate the finite contributions coming from explicit $\epsilon$ or $\epsilon^2$ terms from the BMHV algebra. 

In order to compute the finite BRST violating contributions of every diagram, we retain, in $\mathcal{Z}$, not only the pole, but also those $\epsilon / \epsilon$ contributions that arise from evanescent structures in numerators. To disentangle those from other finite terms, we keep the evanescent numerator dependence, coming from contractions of evanescent metric tensors $\widehat{g}_{\mu\nu}\widehat{g}^{\mu\nu}$ or $\widehat{\gamma}^{\mu}\widehat{\gamma}_{\mu}$ separate from the explicit ultraviolet poles during pole extraction. We denote the corresponding numerator book-keeping parameter by $\eta$, i.e. we set 
\begin{equation}
    \widehat{g}_{\mu\nu}\widehat{g}^{\mu\nu} = -2\eta
\end{equation}
with $\eta \to \epsilon$ imposed only after the $\mathcal{Z}$ operator. 

The bookkeeping parameter $\eta$ tracks dimension factors arising
from evanescent numerator contractions. It does not replace the
$\epsilon$ dependence of tensor-reduction coefficients or scalar
integrals, which are evaluated with $d=4-2\epsilon$. Their products
are expanded in $\epsilon$ before pole extraction, while $\eta$
is kept independent.

At two loops, the $\epsilon$ dependence of these coefficients
can modify both the double-pole and single-pole contributions.
For example, an $O(\epsilon)$ correction multiplying
$\eta/\epsilon^2$ produces a term proportional to
$\eta/\epsilon$, which is retained by pole extraction and becomes
finite when $\eta=\epsilon$ is imposed. Similarly,
$\eta^2/\epsilon^2$ produces a finite contribution.

More generally, after expanding and collecting all coefficient
functions, a retained term $\eta^r\epsilon^{-k}$ becomes
$\epsilon^{r-k}$. It is finite for $r=k$, remains divergent for
$r<k$, and vanishes for $r>k$. The expansions must therefore retain
all orders that can contribute to the extracted poles, accounting
also for poles introduced by reduction. Terms already finite at
fixed $\eta$ are not retained by this UV pole-extraction procedure.

Note that the pole-extraction operator ignores $\eta$:
\begin{equation}
K_\epsilon
\left[
\sum_{n=-N}^{\infty}a_n(\eta)\epsilon^n
\right]
=
\sum_{n=-N}^{-1}a_n(\eta)\epsilon^n.
\end{equation}
Consequently, pole extraction and the identification
$\eta\to \epsilon$ do not commute, as has been noted repeatedly in the literature, see e.g.\cite{Herzog:2017bjx,Caswell:1981ek,Speer:1978dd}. For example,
\begin{equation}
\left.
K_\epsilon\left(\frac{\eta}{\epsilon}\right)
\right|_{\eta=\epsilon}
=1,
\qquad
K_\epsilon
\left[
\left.\frac{\eta}{\epsilon}\right|_{\eta=\epsilon}
\right]
=0.
\end{equation}
We also note that hatted structures involving internal (loop) momenta must not be discarded before the remaining loop integrations have been evaluated.

Operationally, we treat the Lorentz and Dirac algebra as follows: the loop-momentum integrations are performed entirely in $d$ dimensions, while the distinction between the four-dimensional physical subspace and the evanescent complement is retained explicitly through barred and hatted Lorentz and Dirac structures in numerators. 

The loop-momentum dependence is first expressed in terms of tensor integrals that are subsequently reduced (in d-dimensions) to scalar massive vacuum integrals. Integration-by-parts identities reduce these integrals to the
massive tadpole at one loop and, at two loops, to the equal-mass
sunset together with products of one-loop tadpoles. The resulting metric and external momentum tensors are then contracted with the corresponding Dirac and Lorentz structures, using the metric tensors and orthogonality relations of the respective subspaces. Thus, factors originating from the physical and evanescent subspaces remain explicit. Individual diagrams were generated by QGRAF~\cite{Nogueira:1991ex}, and manipulated by in-house routines written in FORM~\cite{Ruijl:2017dtg,Vermaseren:2000nd}. The handling of sub-diagrams and the Taylor (hard region) expansion is performed with a variant of the RQFT library~\cite{benruijl_rqft_2022} written in FORM. 

For the treatment of the Dirac algebra, each $d$-dimensional gamma matrix is decomposed into its four-dimensional and evanescent components. The resulting gamma strings are brought into a normal form in which all evanescent gamma matrices are moved to the left, all four-dimensional gamma matrices to their right, and the chiral projectors $\mathbb{P}_{L/R}$ to the far right. 

Once this normal form has been obtained, Dirac traces can be evaluated algorithmically using the factorization identity
\begin{equation}
\operatorname{tr}\left(
\widehat{\gamma}^{\mu_1}\cdots
\widehat{\gamma}^{\mu_n}
\overline{\gamma}^{\nu_1}\cdots
\overline{\gamma}^{\nu_m}
\boldsymbol{\Lambda}
\right)
=
\frac{1}{4}
\operatorname{tr}\left(
\widehat{\gamma}^{\mu_1}\cdots
\widehat{\gamma}^{\mu_n}
\right)
\operatorname{tr}\left(
\overline{\gamma}^{\nu_1}\cdots
\overline{\gamma}^{\nu_m}
\boldsymbol{\Lambda}
\right),
\end{equation}
where
$\boldsymbol{\Lambda}\in
\{\mathbf{1},\mathbb{P}_L,\mathbb{P}_R\}$
and $\operatorname{tr}(\mathbf{1})=4$.
This factorization identity was proven in
Ref.~\cite{vonManteuffel:2025swv} and allows the evanescent
and four-dimensional parts of the Dirac trace to be treated
independently.

\section{One-loop analysis} 
At the one-loop level, the bare effective action, $\Gamma^{(1)}$, computed with BMHV rules, is divergent and 
BRST breaking. We therefore need to introduce not only UV-singular counter-terms, $S_{\rm sct}^{(1)}$, that render it finite, but also finite counter-terms, $S^{(1)}_{\rm fct}$, that restore BRST symmetry. The renormalized one-loop action becomes

\begin{equation}
\label{eqn:eff_action_one}
\Gamma_{\rm ren}^{(1)}
=S_0+
\Gamma^{(1)}
+S_{\rm sct}^{(1)}
+S_{\rm fct}^{(1)}.
\end{equation}
and the condition for restoration is 
\begin{equation}
\label{eqn:BRST_inv_eff_action}
\lim_{d\rightarrow 4}
\left[
\mathcal{S}_d
\left(
S_0
+
\Gamma^{(1)}
+
S_{\rm sct}^{(1)}
+
S_{\rm fct}^{(1)}
\right)
\right]_{\rm 1-loop}
=
0.
\end{equation}

A direct approach to the restoration of BRST symmetry would consist in computing the complete one-loop effective action, evaluating its Slavnov--Taylor variation, and subsequently identifying the symmetry-breaking contributions from which the necessary finite counter-terms could be reconstructed. Although possible at one loop, such a procedure becomes increasingly cumbersome at higher perturbative orders. Instead, we make use of the Quantum Action Principle (QAP), which provides a systematic and more efficient framework by relating the breaking of the Slavnov--Taylor identity to insertions of local composite operators.

According to the QAP, the breaking of the Slavnov--Taylor identity for the regularized effective action can be represented by the insertion of a local operator. The QAP then yields
\begin{equation}
\mathcal{S}_d\left(\Gamma\right)
=
\Delta\cdot\Gamma,
\end{equation}
where $\Delta\cdot\Gamma$  denotes 1PI Green functions with insertions of a local operator $\Delta$,
which admits a perturbative expansion, $\Delta = \Delta^{(0)}+\Delta^{(1)}+\ldots$.  Its lowest-order contribution is precisely the tree-level evanescent breaking,
\begin{equation}
\Delta^{(0)}
=
\widehat{\Delta}
=
\mathcal{S}_d(S_0)=
\int d^{d}x\,
g_s T^a_{\mathcal R,ij} c^a
\left\{
\overline{\psi}_{i}
\overleftarrow{\widehat{\slashed{\partial}}}
\mathbb{P}_{R}\psi_{j}
+
\overline{\psi}_{i}\mathbb{P}_{L}
\overrightarrow{\widehat{\slashed{\partial}}}
\psi_{j}
\right\}.
\end{equation}
At one-loop order, the relevant breaking is therefore obtained from one-loop Green functions containing a single insertion of $\widehat{\Delta}$. This seemingly trivial statement acquires highly non-trivial content once we note that the QAP asserts its validity also after renormalization.

The corresponding momentum-space insertion is represented by the effective vertex

\begin{equation}
\label{eqn:Delta_hat}
\raisebox{-0.44\height}{
  \includegraphics[
    width=0.3\linewidth
  ]{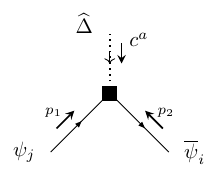}
}
\quad = g_s T_{\mathcal R,ij}^a
\left(
\widehat{\slashed{p}}_{1}\mathbb{P}_{L}
+
\widehat{\slashed{p}}_{2}\mathbb{P}_{R}
\right).
\end{equation}

Extracting the one-loop contribution of the QAP relation gives
\begin{equation}
\label{eqn:QAprinciple}
\left[
\mathcal{S}_d
\left(
S_0+\Gamma^{(1)}
\right)
\right]_{\rm 1-loop}
=
\left(
\widehat{\Delta}\cdot\Gamma^{(1)}
\right).
\end{equation}
where the notation $[\ldots]_{\rm 1-loop}$ indicates that only terms of one-loop order are retained in the expansion of the nonlinear Slavnov--Taylor functional.
\par

Our strategy to compute the restoring counter-terms is therefore to first compute 1PI Green functions containing a single insertion of $\widehat{\Delta}$ whose field content and power-counting degree allow them to generate local operators of canonical dimension not larger than four and ghost number one. This provides us with the BRST violation at one loop order. We, then, have to determine $S_{\rm sct}^{(1)}$ and $S_{\rm fct}^{(1)}$ which will satisfy  Eq.~\ref{eqn:BRST_inv_eff_action}.  

It is easy to see that, retaining only contributions of one-loop order, we can 
write the BRST variation of the complete one loop effective action as 

\begin{equation}
\label{eq:SdManipulationsOneLoop}
\begin{aligned}
\mathcal{S}_d
\left(
S_0
+
\Gamma^{(1)}
+
S_{\rm sct}^{(1)}
+
S_{\rm fct}^{(1)}
\right)_{\rm 1-loop} 
&=
\mathcal{S}_d
\left(
S_0 + \Gamma^{(1)}
\right)_{\rm 1-loop}
+
\mathcal{S}_d
\left(
S_0
+
S_{\rm sct}^{(1)}
+
S_{\rm fct}^{(1)}
\right)_{\rm 1-loop}
\\
&=
\left(\widehat{\Delta}\cdot\Gamma\right)^{(1)}
+
\mathcal S_d\left(
S_0+S_{\rm sct}^{(1)}+S_{\rm fct}^{(1)}
\right)_{\rm 1-loop}
.
\end{aligned}
\end{equation}

The condition for symmetry restoration (eq.~\ref{eqn:BRST_inv_eff_action}) is, therefore, that the four-dimensional limit of the rhs of eq.~\ref{eq:SdManipulationsOneLoop} vanishes. Note that there can still be an evanescent contribution to eq.~\ref{eq:SdManipulationsOneLoop} which is non-zero and will play a role at two loop level.

For the singular part, the one-loop restoration condition takes the form
\begin{equation}
\label{eqn:1-loop-res_div}
\left(
\widehat{\Delta}\cdot\Gamma^{(1)}
\right)_{\rm div}
+
\left[
\mathcal{S}_d
\left(
S_0+S_{\rm sct}^{(1)}
\right)
\right]_{\rm 1-loop}
=
0.
\end{equation}

Equation~\eqref{eqn:1-loop-res_div} provides a consistency check: the insertion
calculation in Sec.~\ref{section:brst-breaking} determines the divergent breaking,
while the counter-terms computed directly in Sec.~\ref{section:one-loop-singular-counterterms}
determine its cancellation through the Slavnov--Taylor
identity application.

The corresponding restoration condition for the finite part is
\begin{equation}
\label{eqn:1-loop-res_fin}
\lim_{d\rightarrow4}
\left\{
\left(
\widehat{\Delta}\cdot\Gamma^{(1)}
\right)_{\rm fin}
+
\left[
\mathcal{S}_d
\left(
S_0+S_{\rm fct}^{(1)}
\right)
\right]_{\rm 1-loop}
\right\}
=
0.
\end{equation}
We determine $S_{\rm fct}^{(1)}$ by constructing a local functional
whose linearized Slavnov--Taylor variation cancels the finite
breaking.
\par
We first present the singular counter-terms obtained directly
from ordinary one-loop 1PI functions in Sec.~\ref{section:one-loop-singular-counterterms}. We then compute
the breaking from $\widehat{\Delta}$ insertions in Sec.~\ref{section:brst-breaking} and
construct the finite symmetry-restoring counter-terms in Sec.~\ref{sec:Derivation of finite counterterms}.

\subsection{One loop singular counter-terms}
\label{section:one-loop-singular-counterterms}
In this section, we present the singular counter-terms for
the one-loop 1PI Green functions evaluated in our analysis. At one loop, we retain a general covariant-gauge parameter $\xi$.
The Feynman-gauge results are recovered by setting $\xi=1$.
\par
In the expressions below, $S_{\mathrm{mom}}^{\mathcal{O},(1)}$ denotes the momentum-space
Feynman rule corresponding to the operator $\mathcal{O}$, while $S_{\mathrm{sct}}^{\mathcal{O},(1)}$ denotes the corresponding integrated local
counter-term. The field ordering shown in the local expressions
is important when treating Grassmann-valued fields. Combinatorial factors
for identical fields are included in the local functionals.

\begin{itemize}[leftmargin=0pt]
\item 
\textbf{Fermionic self-energy}:
\quad
\raisebox{-0.4\height}{
  \includegraphics[
    width=0.22\linewidth
  ]{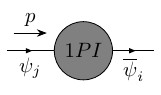}
}
\begin{equation}
\left(S_{\text{sct,mom}}^{\psi\overline{\psi},(1)}\right)
=
-\frac{g_s^2}{(4\pi)^2}
\frac{\xi}{\epsilon}
\left[C_2(\mathcal R)\right]_{ij}
\overline{\slashed p}\,\mathbb P_R,
\end{equation}
\begin{equation}
\left(S_{\text{sct}}^{\psi\overline{\psi},(1)}\right)=
-i\,\frac{g_s^2}{(4\pi)^2}
\frac{\xi}{\epsilon}
\int d^d x\,
\overline{\psi}_i\,
\left[C_2(\mathcal R)\right]_{ij}
\overline{\slashed{\partial}}\,
\mathbb P_R\,\psi_j .
\end{equation}

\item 
\textbf{Gauge boson self-energy}:
\quad
\raisebox{-0.4\height}{
  \includegraphics[
    width=0.22\linewidth
  ]{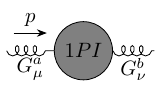}
}
\begin{equation}
\begin{aligned}
\left(S_{\text{sct,mom}}^{GG,(1)}\right)
&=
\frac{g_s^2}{(4\pi)^2}
\frac{\delta^{ab}}{\epsilon}
\Bigg\{
\frac{13-3\xi}{6}C_2(G)
\left(p^\mu p^\nu-p^2g^{\mu\nu}\right) \\
&\qquad
+\frac{2}{3}\sum_m C(R_m)
\left[
-\overline p^\mu\overline p^\nu
+\overline p^{\,2}\overline g^{\mu\nu}
+\frac12\widehat p^{\,2}\overline g^{\mu\nu}
\right]
\Bigg\}
\end{aligned}
\end{equation}
\begin{equation}
\begin{aligned}
\left(S_{\text{sct}}^{GG,(1)}\right)&=
\frac{g_s^2}{2(4\pi)^2\epsilon}
\int d^d x\,G_\mu^a
\Bigg\{
\frac{13-3\xi}{6}C_2(G)
\left(
g^{\mu\nu}\partial^2
-\partial^\mu\partial^\nu
\right)
\\[0.5em]
&\qquad
+\frac{2}{3}\sum_m C(R_m)
\left[
\overline\partial^\mu\overline\partial^\nu
-\overline\partial^{\,2}\overline g^{\mu\nu}
-\frac12\widehat\partial^{\,2}\overline g^{\mu\nu}
\right]
\Bigg\}
G_\nu^a .
\end{aligned}
\end{equation}

\item 
\textbf{Ghost self-energy}:
\quad
\raisebox{-0.4\height}{
  \includegraphics[
    width=0.22\linewidth
  ]{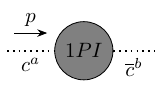}
}

\begin{equation}
\begin{aligned}
\left(S_{\text{sct,mom}}^{c\overline c,(1)}\right)
&=
\frac{g_s^2}{(4\pi)^2}
\delta^{ab}C_2(G)\,
\frac{3-\xi}{4\epsilon}\,p^2,
\end{aligned}
\end{equation}
\begin{equation}
\begin{aligned}
\left(S_{\text{sct}}^{c\overline c,(1)}\right)&=
-\frac{g_s^2}{(4\pi)^2}
C_2(G)\,\frac{3-\xi}{4\epsilon}
\int d^d x\,
\overline c^{\,a}\,\partial^2 c^a .
\end{aligned}
\end{equation}

\item 
\textbf{Ghost-BRST-source function}:
\quad
\raisebox{-0.4\height}{
  \includegraphics[
    width=0.22\linewidth
  ]{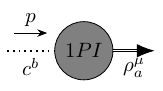}
}

\begin{equation}
\begin{aligned}
\left(S_{\text{sct,mom}}^{\rho c,(1)}\right)
&=
-\frac{i g_s^2}{(4\pi)^2}
C_2(G)\frac{3-\xi}{4\epsilon}
\delta^{ab}p^\mu,
\end{aligned}
\end{equation}
\begin{equation}
\begin{aligned}
\left(S_{\text{sct}}^{\rho c,(1)}\right)&=
\frac{g_s^2}{(4\pi)^2}
C_2(G)\frac{3-\xi}{4\epsilon}
\int d^d x\,
\rho_\mu^a\,\partial^\mu c^a .
\end{aligned}
\end{equation}

\item 
\textbf{Fermion-gauge boson vertex}:
\quad
\raisebox{-0.4\height}{
  \includegraphics[
    width=0.3\linewidth
  ]{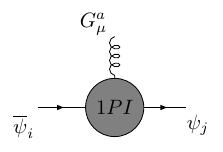}
}

\begin{equation}
\begin{aligned}
\left(S_{\text{sct,mom}}^{\psi\overline{\psi}G,(1)}\right)
&=
-\frac{g_s^3}{(4\pi)^2\epsilon}
\left[
T_{\mathcal R}^a
\left(
\frac{3+\xi}{4}C_2(G)\mathbf 1
+\xi C_2(\mathcal R)
\right)
\right]_{ij}
\overline{\gamma}^{\mu}\mathbb P_R,
\end{aligned}
\end{equation}
\begin{equation}
\begin{aligned}
\left(S_{\text{sct}}^{\psi\overline{\psi}G,(1)}\right)&=
-\frac{g_s^3}{(4\pi)^2\epsilon}
\int d^d x\,
\overline{\psi}_i\,\overline{\gamma}^{\mu}
\\[0.3em]
&\qquad\times
\left[
T_{\mathcal R}^a
\left(
\frac{3+\xi}{4}C_2(G)\mathbf 1
+\xi C_2(\mathcal R)
\right)
\right]_{ij}
\mathbb P_R\,\psi_j\,G_\mu^a .
\end{aligned}
\end{equation}

\item 
\textbf{Ghost-gauge boson vertex}:
\quad
\raisebox{-0.4\height}{
  \includegraphics[
    width=0.3\linewidth
  ]{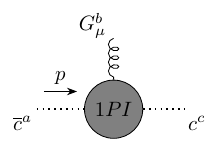}
}

\begin{equation}
\begin{aligned}
\left(S_{\text{sct,mom}}^{c\overline cG,(1)}\right)
&=
\frac{i g_s^3}{(4\pi)^2}
f^{abc}C_2(G)\frac{\xi}{2\epsilon}\,
p^\mu,
\end{aligned}
\end{equation}
\begin{equation}
\begin{aligned}
\left(S_{\text{sct}}^{c\overline cG,(1)}\right)&=
-\frac{g_s^3}{(4\pi)^2}
C_2(G)\frac{\xi}{2\epsilon}
\int d^d x\,
f^{abc}\,
\left(\partial^\mu\overline c^{\,a}\right)
G_\mu^b\,c^c .
\end{aligned}
\end{equation}

\item 
\textbf{Triple-gauge boson vertex}:
\quad
\raisebox{-0.4\height}{
  \includegraphics[
    width=0.3\linewidth
  ]{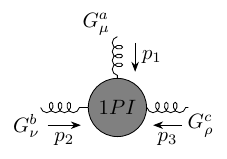}
}

\begin{equation}
\begin{aligned}
\left(S_{\text{sct,mom}}^{GGG,(1)}\right)
&=
\frac{i g_s^3}{(4\pi)^2\epsilon}
f^{abc}
\Bigg\{
\frac{2}{3}\sum_m C(R_m)
\Big[
(\overline p_2-\overline p_3)^\mu
\overline g^{\nu\rho}
+(\overline p_3-\overline p_1)^\nu
\overline g^{\mu\rho}
+(\overline p_1-\overline p_2)^\rho
\overline g^{\mu\nu}
\Big]
\\[0.5em]
&\qquad
-\frac{17-9\xi}{12}C_2(G)
\Big[
(p_2-p_3)^\mu g^{\nu\rho}
+(p_3-p_1)^\nu g^{\mu\rho}
+(p_1-p_2)^\rho g^{\mu\nu}
\Big]
\Bigg\},
\end{aligned}
\end{equation}

\begin{equation}
\begin{aligned}
\left(S_{\text{sct}}^{GGG,(1)}\right)&=
\frac{g_s^3}{(4\pi)^2\epsilon}
\int d^d x\,f^{abc}
\Bigg\{
\frac{2}{3}\sum_m C(R_m)
\left(\overline\partial_\mu\overline G_\nu^a\right)
\overline G^{b\mu}\overline G^{c\nu}
 \\[0.5em]
 &\qquad
-\frac{17-9\xi}{12}C_2(G)
\left(\partial_\mu G_\nu^a\right)
G^{b\mu}G^{c\nu}
\Bigg\}.
\end{aligned}
\end{equation}

\item 
\textbf{Ghost-gluon-BRST-source vertex}:
\quad
\raisebox{-0.4\height}{
  \includegraphics[
    width=0.3\linewidth
  ]{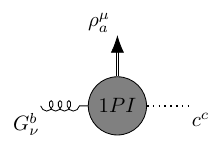}
}

\begin{equation}
\begin{aligned}
\left(S_{\text{sct,mom}}^{\rho cG,(1)}\right)
&=
-\frac{g_s^3}{(4\pi)^2}
C_2(G)\frac{\xi}{2\epsilon}\,
g^{\mu\nu}f^{abc},
\end{aligned}
\end{equation}
\begin{equation}
\begin{aligned}
\left(S_{\text{sct}}^{\rho cG,(1)}\right)&=
-\frac{g_s^3}{(4\pi)^2}
C_2(G)\frac{\xi}{2\epsilon}
\int d^d x\,
f^{abc}\,
\rho_\mu^a\,G^{b\mu}\,c^c .
\end{aligned}
\end{equation}

\item 
\textbf{Ghost-ghost-BRST-source vertex}:
\quad
\raisebox{-0.4\height}{
  \includegraphics[
    width=0.3\linewidth
  ]{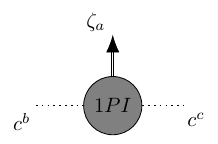}
}

\begin{equation}
\begin{aligned}
\left(S_{\text{sct,mom}}^{\zeta cc,(1)}\right)
&=
\frac{g_s^3}{(4\pi)^2}
C_2(G)\frac{\xi}{2\epsilon}\,
f^{abc},
\end{aligned}
\end{equation}
\begin{equation}
\begin{aligned}
\left(S_{\text{sct}}^{\zeta cc,(1)}\right)&=
\frac{g_s^3}{(4\pi)^2}
C_2(G)\frac{\xi}{2\epsilon}
\int d^d x\,
\frac12 f^{abc}\,
\zeta^a\,c^b c^c .
\end{aligned}
\end{equation}

\item 
\textbf{Fermion-ghost-BRST-source vertex}:
\quad
\raisebox{-0.4\height}{
  \includegraphics[
    width=0.3\linewidth
  ]{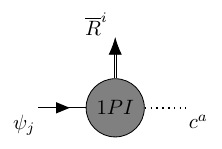}
}

\begin{equation}
\begin{aligned}
\left(S_{\text{sct,mom}}^{\overline R c\psi,(1)}\right)
&=
-\frac{i g_s^3}{(4\pi)^2}
T_{\mathcal R,ij}^a C_2(G)
\frac{\xi}{2\epsilon}\,\mathbb P_R,
\end{aligned}
\end{equation}
\begin{equation}
\begin{aligned}
\left(S_{\text{sct}}^{\overline R c\psi,(1)}\right)&=
-\frac{i g_s^3}{(4\pi)^2}
C_2(G)\frac{\xi}{2\epsilon}
\int d^d x\,
\overline R^{\,i}\,
T_{\mathcal R,ij}^a\,
c^a\,\mathbb P_R\,\psi^j .
\end{aligned}
\end{equation}

\item
\textbf{Anti-Fermion-ghost-BRST-source vertex}:
\quad
\raisebox{-0.4\height}{
  \includegraphics[
    width=0.3\linewidth
  ]{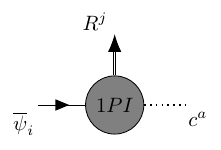}
}

\begin{equation}
\begin{aligned}
\left(S_{\text{sct,mom}}^{Rc\overline\psi,(1)}\right)
&=
-\frac{i g_s^3}{(4\pi)^2}
T_{\mathcal R,ij}^a C_2(G)
\frac{\xi}{2\epsilon}\,\mathbb P_L,
\end{aligned}
\end{equation}
\begin{equation}
\begin{aligned}
\left(S_{\text{sct}}^{Rc\overline\psi,(1)}\right)&=
-\frac{i g_s^3}{(4\pi)^2}
C_2(G)\frac{\xi}{2\epsilon}
\int d^d x\,
\overline\psi^{\,i}\,
\mathbb P_L\,
T_{\mathcal R,ij}^a\,
c^a\,R^j .
\end{aligned}
\end{equation}

\item
\textbf{Four-gluon vertex}:
\quad
\raisebox{-0.4\height}{
  \includegraphics[
    width=0.3\linewidth
  ]{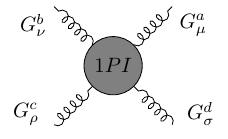}
}

\begin{equation}
\begin{aligned}
\left(S_{\mathrm{sct,mom}}^{GGGG,(1)}\right)
&=
\frac{g_s^4}{3(4\pi)^2\epsilon}
\Bigg[
(3\xi-2)C_2(G)
\mathcal{F}^{abcd}_{\mu\nu\rho\sigma}
+2\sum_m C(R_m)
\overline{\mathcal{F}}^{abcd}_{\mu\nu\rho\sigma}
\Bigg],
\end{aligned}
\end{equation}
where 
\begin{equation}
    \mathcal{F}^{abcd}_{\mu\nu\rho\sigma} = \left(
g_{\mu\nu}g_{\rho\sigma},\,
g_{\mu\rho}g_{\nu\sigma},\,
g_{\mu\sigma}g_{\nu\rho}
\right)
\begin{pmatrix}
f^{ace}f^{bde}+f^{ade}f^{bce}\\
f^{abe}f^{cde}+f^{ade}f^{cbe}\\
f^{abe}f^{dce}+f^{ace}f^{dbe}
\end{pmatrix}
\end{equation}
\begin{equation}
    \overline{\mathcal{F}}^{abcd}_{\mu\nu\rho\sigma} = \left(
\bar g_{\mu\nu}\bar g_{\rho\sigma},\,
\bar g_{\mu\rho}\bar g_{\nu\sigma},\,
\bar g_{\mu\sigma}\bar g_{\nu\rho}
\right)
\begin{pmatrix}
f^{ace}f^{bde}+f^{ade}f^{bce}\\
f^{abe}f^{cde}+f^{ade}f^{cbe}\\
f^{abe}f^{dce}+f^{ace}f^{dbe}
\end{pmatrix}
\end{equation}

\begin{equation}
\begin{aligned}
\left(S_{\mathrm{sct}}^{GGGG,(1)}\right)&=
\frac{g_s^4}{12(4\pi)^2\epsilon}
\int d^d x\,f^{abe}f^{cde}
\Bigg[
(3\xi-2)C_2(G)\,
G_\mu^a G_\nu^b G^{c\mu}G^{d\nu}
 \\[0.5em]
 &\qquad
+2\sum_m C(R_m)\,
\bar G_\mu^a\bar G_\nu^b
\bar G^{c\mu}\bar G^{d\nu}
\Bigg].
\end{aligned}
\end{equation}
\end{itemize}
All one-loop divergent 1PI functions presented
above agree with those of Ref.~\cite{BeluscaMaito_2020}, upon
specializing its model, which also includes scalars and Yukawa
interactions, to our theory without scalar fields. Agreement is also confirmed
with the singular counterterms of
Ref.~\cite{Martin:1999cc}, which considers a scalar-free
theory with both left- and right-handed fermions, after restricting
its gauge group to $SU(N)$ and retaining only the right-handed
fermionic sector. In the latter comparison, we account for the
different field normalizations and the convention $\epsilon=4-d$
used there. Both comparisons include the evanescent divergent terms.

\subsection{BRST violating terms at one loop}
\label{section:brst-breaking}

We consider 1PI Green functions containing a single
$\widehat{\Delta}$ insertion whose local breaking has canonical
dimension at most four and ghost number one. We assign the canonical dimensions
$[G]=1$, $[\psi]=[\overline\psi]=3/2$,
$[c]=0$, $[\overline c]=2$, $[\rho]=3$,
$[R]=[\overline R]=5/2$, and $[\zeta]=4$. For a given channel,
the maximum momentum degree of a local contribution is
\begin{equation}
\omega=4-\sum_{\Phi}[\Phi],
\end{equation}
where the sum runs over the external fields. 

The candidate channels compatible with power counting,
ghost number one, and fermion-number conservation are listed
in Table~\ref{tab:brst-breaking-channels}. The first eight admit
one-loop 1PI diagrams, whereas the remaining five first admit
such diagrams at two loops. The existence of diagrams does
not imply a nonzero local breaking.

\begin{table}[htbp]
\centering
\small
\setlength{\tabcolsep}{5pt}
\begin{tabular}{lccc}
\hline
\shortstack{External-field\\channel}
&
\shortstack{Field dimension\\$\sum_{\Phi}[\Phi]$}
&
\shortstack{Momentum\\degree $\omega$}
&
\shortstack{First perturbative\\loop order}
\\
\hline
$Gc$                         & 1 & 3 & 1 \\
$GGc$                        & 2 & 2 & 1 \\
$GGGc$                       & 3 & 1 & 1 \\
$GGGGc$                      & 4 & 0 & 1 \\
$\psi\overline\psi c$         & 3 & 1 & 1 \\
$\psi\overline\psi Gc$        & 4 & 0 & 1 \\
$\overline R\psi cc$          & 4 & 0 & 1 \\
$R\overline\psi cc$           & 4 & 0 & 1 \\
\hline
$\overline c\,cc$             & 2 & 2 & 2 \\
$\overline c\,Gcc$            & 3 & 1 & 2 \\
$\rho\,cc$                   & 3 & 1 & 2 \\
$\rho\,Gcc$                  & 4 & 0 & 2 \\
$\zeta\,ccc$                 & 4 & 0 & 2 \\
\hline
\end{tabular}
\caption{Candidate 1PI channels with a BRST-breaking insertion,
for local contributions of canonical dimension at most four
and ghost number one. The dimension shown counts the external
fields and sources, without including derivatives. The last
column gives the first perturbative loop order, counting both
the integration loops and the perturbative order of the
vertices.}
\label{tab:brst-breaking-channels}
\end{table}

Each external antighost supplies at least one external momentum.
This excludes local contributions in the channels
$GG\overline c\,cc$ and $\overline c\,\overline c\,ccc$,
for which $\omega=0$. A term containing only one ghost cannot form a colour singlet,
because there is no other colour index to contract with its
index. The linear field and source equations exclude additional
radiative 1PI channels involving $B$ or $\chi$.

In this section, we present the resulting one-loop contributions
$\left(\widehat{\Delta}\cdot\Gamma^{(1)}\right)_i$,
distinguishing their singular and finite parts. The index $i$
denotes the corresponding external-field channel.

\begin{itemize}[leftmargin=0pt]
\item 
\textbf{Gauge-boson-ghost function}:
\quad
\raisebox{-0.4\height}{
  \includegraphics[
    width=0.22\linewidth
  ]{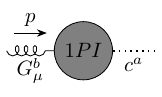}
}

\begin{equation}
\begin{aligned}
\label{eqn:Gc_viol_1_loop}
\left(\widehat{\Delta}\cdot\Gamma^{(1)}\right)_{Gc,\text{mom}}
&=
-\frac{i g_s^2}{3(4\pi)^2}
\sum_m C(R_m)\,\delta^{ab}
\left(
\overline{p}^{\mu}\widehat{p}^{\,2}\frac{1}{\epsilon}
+\overline{p}^{\mu}p^2
\right),
\end{aligned}
\end{equation}
\begin{equation}
\begin{aligned}
\left(\widehat{\Delta}\cdot\Gamma^{(1)}\right)_{Gc}&=
\frac{g_s^2}{3(4\pi)^2}
\sum_m C(R_m)
\int d^d x\,
\left(
\frac{1}{\epsilon}\,
\overline{\partial}_{\mu}c^a\,
\widehat{\partial}^{\,2}G^{\mu,a}
+
\overline{\partial}_{\mu}c^a\,
\partial^2G^{\mu,a}
\right).
\end{aligned}
\end{equation}

\item 
\textbf{Double-gauge-boson-ghost vertex}:
\quad
\raisebox{-0.4\height}{
  \includegraphics[
    width=0.3\linewidth
  ]{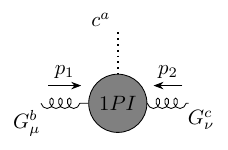}
}
\begin{equation}
\begin{aligned}
\left(\widehat{\Delta}\cdot\Gamma^{(1)}\right)_{GGc,\text{mom}}
&= -\frac{g_s^3}{3(4\pi)^2}\sum_m C(R_m)\,f^{abc}
 \\
 &\quad\times
\Biggl[
\frac{1}{\epsilon}
\left(\widehat{p}_1^2-\widehat{p}_2^2\right)
\overline{g}^{\mu\nu}
-\left(\overline{p}_1^2-\overline{p}_2^2\right)
\overline{g}^{\mu\nu}
+2\overline{p}_1^\mu\overline{p}_1^\nu
-2\overline{p}_2^\mu\overline{p}_2^\nu
\Biggr],
\end{aligned}
\end{equation}

\begin{equation}
\begin{aligned}
\left(\widehat{\Delta}\cdot\Gamma^{(1)}\right)_{GGc}
&= -\frac{g_s^3}{3(4\pi)^2}\sum_m C(R_m)\,f^{abc}
 \\
 &\quad\times
\int d^d x\,c^a\Biggl[
-\frac{1}{\epsilon}\,
\overline{G}_\mu^c\widehat{\partial}^2
\overline{G}^{\mu,b}
+\overline{G}_\mu^c\overline{\partial}^2
\overline{G}^{\mu,b}
-2\overline{G}^{\nu,c}
\overline{\partial}_\nu\overline{\partial}_\mu
\overline{G}^{\mu,b}
\Biggr].
\end{aligned}
\end{equation}

\item 
\textbf{Triple-gauge-boson-ghost vertex}:
\quad
\raisebox{-0.4\height}{
  \includegraphics[
    width=0.3\linewidth
  ]{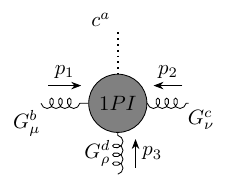}
}

\begin{equation}
\begin{aligned}
\left(\widehat{\Delta}\cdot\Gamma^{(1)}\right)_{GGGc,\text{mom}}
&=
-\frac{i g_s^4}{3(4\pi)^2}
\left(\overline{p}_1+\overline{p}_2+\overline{p}_3\right)_\sigma
 \\[0.5em]
 &\quad\times
\left[
\mathcal{A}_{\mathcal R}^{abcd}\,
\overline{g}^{\mu\nu}\overline{g}^{\rho\sigma}
+\mathcal{A}_{\mathcal R}^{abdc}\,
\overline{g}^{\mu\rho}\overline{g}^{\nu\sigma}
+\mathcal{A}_{\mathcal R}^{acdb}\,
\overline{g}^{\mu\sigma}\overline{g}^{\nu\rho}
\right],
\end{aligned}
\end{equation}
\begin{equation}
\begin{aligned}
\left(\widehat{\Delta}\cdot\Gamma^{(1)}\right)_{GGGc}&=
\frac{g_s^4}{6(4\pi)^2}\mathcal{A}_{\mathcal R}^{abcd}
\int d^d x\,c^a\,\partial_\nu
\left(
\overline{G}_\mu^b
\overline{G}^{\mu,c}
\overline{G}^{\nu,d}
\right),
\end{aligned}
\end{equation}
where the ordered and symmetrized quartic traces are those
introduced in the group-theory conventions, and
\begin{equation}
\begin{aligned}
\mathcal{A}_{\mathcal R}^{abcd}
&=
T_{\mathcal R}^{abcd}
+T_{\mathcal R}^{acbd}
+T_{\mathcal R}^{adbc}
+T_{\mathcal R}^{adcb}
-T_{\mathcal R}^{abdc}
-T_{\mathcal R}^{acdb}
\\[0.5em]
&=
2d_{\mathcal R}^{abcd}
-\frac{2}{3}\sum_m C(R_m)
\left(
f^{abe}f^{cde}+f^{ace}f^{bde}
\right)
=
\sum_m \mathcal{A}_{R_m}^{abcd}.
\end{aligned}
\end{equation}

\item
\textbf{Fermion-antifermion-ghost vertex}:
\quad
\raisebox{-0.3\height}{
  \includegraphics[
    width=0.3\linewidth
  ]{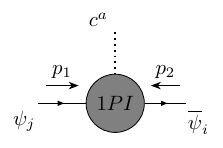}
}

\begin{equation}
\begin{aligned}
\left(\widehat{\Delta}\cdot\Gamma^{(1)}\right)_{\psi\overline{\psi}c,\text{mom}}
&=
-\frac{i g_s^3}{(4\pi)^2}
\Bigg\{
\frac{\xi C_2(G)}{2\epsilon}\,
T_{\mathcal R,ij}^{a}
\left(
\widehat{\slashed p}_1\mathbb P_L
+\widehat{\slashed p}_2\mathbb P_R
\right)
\\[0.5em]
&\qquad
-\left[
T_{\mathcal R}^{a}
\left(
\frac{5+\xi}{6}C_2(\mathcal R)
-\frac{\xi}{4}C_2(G)\mathbf 1
\right)
\right]_{ij}
\left(
\overline{\slashed p}_1+\overline{\slashed p}_2
\right)\mathbb P_R
\Bigg\},
\end{aligned}
\end{equation}
\begin{equation}
\begin{aligned}
\left(\widehat{\Delta}\cdot\Gamma^{(1)}\right)_{\psi\overline{\psi}c}&=
\frac{g_s^3}{(4\pi)^2}
\int d^d x\,c^a
\Bigg\{
\frac{\xi C_2(G)}{2\epsilon}\,
T_{\mathcal R,ij}^{a}
\left(
\overline\psi_i
\overleftarrow{\widehat{\slashed{\partial}}}
\mathbb P_R\psi_j
+
\overline\psi_i\mathbb P_L
\overrightarrow{\widehat{\slashed{\partial}}}\psi_j
\right)
\\[0.5em]
&\qquad
-\left[
T_{\mathcal R}^{a}
\left(
\frac{5+\xi}{6}C_2(\mathcal R)
-\frac{\xi}{4}C_2(G)\mathbf 1
\right)
\right]_{ij}
\left(
\overline\psi_i
\overleftarrow{\overline{\slashed{\partial}}}
\mathbb P_R\psi_j
+
\overline\psi_i\mathbb P_L
\overrightarrow{\overline{\slashed{\partial}}}\psi_j
\right)
\Bigg\}.
\end{aligned}
\end{equation}

\item
\textbf{Fermion-antifermion-gauge-boson-ghost vertex}:
\quad
\raisebox{-0.4\height}{
  \includegraphics[
    width=0.3\linewidth
  ]{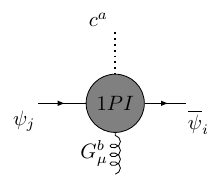}
}
\begin{equation}
\begin{aligned}
\left(\widehat{\Delta}\cdot\Gamma^{(1)}\right)_{\psi\overline{\psi}Gc,\text{mom}}
&=
-\frac{i g_s^4}{(4\pi)^2}
\frac{\xi C_2(G)}{4}
\left[T_{\mathcal R}^a,T_{\mathcal R}^b\right]_{ij}
\overline{\gamma}^{\mu}\mathbb P_R,
\end{aligned}
\end{equation}
\begin{equation}
\begin{aligned}
\left(\widehat{\Delta}\cdot\Gamma^{(1)}\right)_{\psi\overline{\psi}Gc}&=
\frac{g_s^4}{(4\pi)^2}
\frac{\xi C_2(G)}{4}
f^{abc}T_{\mathcal R,ij}^c
\int d^d x\,c^a\,
\overline{\psi}_i\,
\overline{\slashed G}^{\,b}
\mathbb P_R\psi_j .
\end{aligned}
\end{equation}

\item 
\textbf{BRST-source-fermion-double-ghost vertex}:
\quad
\raisebox{-0.5\height}{
  \includegraphics[
    width=0.22\linewidth
  ]{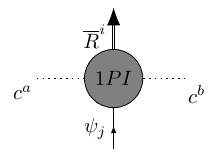}
}
\begin{equation}
\begin{aligned}
\left(\widehat{\Delta}\cdot\Gamma^{(1)}\right)_{\overline R\psi cc,\text{mom}}
&=
\frac{g_s^4}{(4\pi)^2}
\frac{\xi C_2(G)}{4}
\left[T_{\mathcal R}^a,T_{\mathcal R}^b\right]_{ij}
\mathbb P_R,
\end{aligned}
\end{equation}
\begin{equation}
\begin{aligned}
\left(\widehat{\Delta}\cdot\Gamma^{(1)}\right)_{\overline R\psi cc}&=
\frac{i g_s^4}{(4\pi)^2}
\frac{\xi C_2(G)}{8}
f^{abc}T_{\mathcal R,ij}^c
\int d^d x\,
\overline R^{\,i}c^a c^b
\mathbb P_R\psi_j .
\end{aligned}
\end{equation}

\item
\textbf{BRST-source-anti-fermion-double-ghost vertex}:
\quad
\raisebox{-0.5\height}{
  \includegraphics[
    width=0.22\linewidth
  ]{BRST_source_anti_fermion_double_ghost}
}

\begin{equation}
\begin{aligned}
\left(\widehat{\Delta}\cdot\Gamma^{(1)}\right)_{R\overline\psi cc,\text{mom}}
&=
-\frac{g_s^4}{(4\pi)^2}
\frac{\xi C_2(G)}{4}
\left[T_{\mathcal R}^a,T_{\mathcal R}^b\right]_{ij}
\mathbb P_L,
\end{aligned}
\end{equation}
\begin{equation}
\begin{aligned}
\left(\widehat{\Delta}\cdot\Gamma^{(1)}\right)_{R\overline\psi cc}&=
-\frac{i g_s^4}{(4\pi)^2}
\frac{\xi C_2(G)}{8}
f^{abc}T_{\mathcal R,ij}^c
\int d^d x\,
\overline\psi_i\,c^a c^b
\mathbb P_L R^j .
\end{aligned}
\end{equation}

\end{itemize}

\subsection{Finite BRST-restoring counter-terms at one loop}
\label{sec:Derivation of finite counterterms}

Assuming perturbative gauge-anomaly cancellation, the finite
one-loop breaking is cohomologically trivial: it is the
linearized Slavnov--Taylor variation of a local functional.
A finite local counter-term can therefore cancel it.

In the following restoration identities, the local functionals on the right-hand side are understood in the physical four-dimensional limit.

For the gauge-field sectors, we obtain
\begin{equation}
\begin{aligned}
&\lim_{d\rightarrow 4}
\left(\widehat{\Delta}\cdot\Gamma^{(1)}_{\rm fin}\right)_{Gc}
+\lim_{d\rightarrow 4}
\left(\widehat{\Delta}\cdot\Gamma^{(1)}_{\rm fin}\right)_{GGc}
\\[0.5em]
&=
-s_d \left\{
\frac{g_s^2}{3(4\pi)^2}
\sum_m C(R_m)\,\delta^{ab}
\int d^d x\,
\left[
\frac{1}{2}\overline{G}_\mu^a
\overline{\partial}^{\,2}\overline{G}^{\mu,b}
-\overline{G}_\mu^a\overline{\partial}^{\mu}
\overline{\partial}^{\nu}\overline{G}_\nu^b
\right]
\right\}.
\end{aligned}
\end{equation}
and

\begin{equation}
\begin{aligned}
\lim_{d\rightarrow 4}
\left(\widehat{\Delta}\cdot\Gamma^{(1)}_{\rm fin}\right)_{GGGc}
&=
-s_d \left\{
\frac{g_s^4}{12(4\pi)^2}
\left[
T_{\mathcal R}^{abcd}
\int d^d x\,G_\mu^a G^{\mu,b}G_\nu^c G^{\nu,d}
\right.\right.
\\[0.5em]
&\qquad\left.\left.
+\frac{1}{2}\sum_m C(R_m)\,f^{eac}f^{ebd}
\int d^d x\,G_\mu^a G^{\mu,b}G_\nu^c G^{\nu,d}
\right]
\right\}.
\end{aligned}
\end{equation}

The second relation follows from the trace identity
\begin{equation}
\left(\sum_m C(R_m)\right)f^{eac}f^{ebd}
=
T_{\mathcal R}^{abdc}
+T_{\mathcal R}^{acdb}
-T_{\mathcal R}^{acbd}
-T_{\mathcal R}^{adbc},
\end{equation}
together with the symmetry under $b\leftrightarrow c$.

For the fermion and external-source sectors, we obtain
\begin{equation}
\begin{aligned}
&\lim_{d\rightarrow4}
\left(\widehat{\Delta}\cdot\Gamma_{\rm fin}^{(1)}\right)_{\psi\overline\psi c}
+
\lim_{d\rightarrow4}
\left(\widehat{\Delta}\cdot\Gamma_{\rm fin}^{(1)}\right)_{\psi\overline\psi Gc}
 \\[0.5em]
 &\quad
+\lim_{d\rightarrow4}
\left(\widehat{\Delta}\cdot\Gamma_{\rm fin}^{(1)}\right)_{\overline R\psi cc}
+
\lim_{d\rightarrow4}
\left(\widehat{\Delta}\cdot\Gamma_{\rm fin}^{(1)}\right)_{R\overline\psi cc}
\\[0.8em]
&=
-\mathcal S_d\Bigg\{
S_0+\frac{g_s^2}{(4\pi)^2}
\int d^d x\,
\Bigg[
-\frac{\xi C_2(G)}{4}
\left(
\overline R^{\,i}s_d\psi_{Ri}
+R^i s_d\overline\psi_{Ri}
\right)
 \\[0.5em]
 &\qquad
+i\,\frac{5+\xi}{6}\,
\overline\psi_i
\left[C_2(\mathcal R)\right]_{ij}
\overline{\slashed{\partial}}\,
\mathbb P_R\psi_j
\Bigg]
\Bigg\}\Bigg|_{\text{1-loop}}
\\[0.8em]
&=
-\mathcal S_d\Bigg\{
S_0-\frac{g_s^2}{(4\pi)^2}
\frac{\xi C_2(G)}{4}
\int d^d x\,
\left(
\overline R^{\,i}s_d\psi_{Ri}
+R^i s_d\overline\psi_{Ri}
\right)
\Bigg\}\Bigg|_{\text{1-loop}}
\\[0.5em]
&\quad
-s_d\Bigg\{
\frac{i g_s^2}{(4\pi)^2}
\frac{5+\xi}{6}
\int d^d x\,
\overline\psi_i
\left[C_2(\mathcal R)\right]_{ij}
\overline{\slashed{\partial}}\,
\mathbb P_R\psi_j
\Bigg\}.
\end{aligned}
\end{equation}

Here only the one-loop contribution is retained, and the
physical four-dimensional projection is understood. Evanescent
terms are omitted from these projected identities, but must be
retained when needed for insertions into higher-loop diagrams.
We use the following source identity. The added term proportional
to $g_s^2$ is assigned one-loop order and the indicated projection
therefore retains only terms linear in this addition.

\begin{equation}
\label{eqn:R-fct-var}
\begin{aligned}
&\left.
\mathcal S_d
\Bigg\{
S_0
+g_s^2\int d^d x\,
\left(
\overline R^{\,i}s_d\psi_{Ri}
+R^i s_d\overline\psi_{Ri}
\right)
\Bigg\}
\right|_{4D,\,\mathrm{1\text{-}loop}}
\\[0.5em]
&=
g_s^3 T_{\mathcal R,ij}^{a}
\int d^d x\,c^a\,
\overline\partial_\mu
\left(
\overline\psi_i\overline\gamma^\mu
\mathbb P_R\psi_j
\right)
-g_s^4 f^{abc}T_{\mathcal R,ij}^{c}
\int d^d x\,
\overline\psi_i c^a
\overline{\slashed G}^{b}\mathbb P_R\psi_j
\\[0.5em]
&\quad
+\frac{i g_s^4}{2}f^{abc}T_{\mathcal R,ij}^{c}
\int d^d x\,
\overline R^{\,i}c^a c^b\mathbb P_R\psi_j
-\frac{i g_s^4}{2}f^{abc}T_{\mathcal R,ij}^{c}
\int d^d x\,
\overline\psi_i\mathbb P_L c^a c^b R^j .
\end{aligned}
\end{equation}

Combining these relations gives a finite symmetry-restoring
counter-term action satisfying Eq.~\eqref{eqn:1-loop-res_fin}:

\begin{equation}
\begin{aligned}
S_{\rm fct}^{(1)}
&=
\frac{g_s^2}{(4\pi)^2}
\Bigg\{
\frac{1}{6}\sum_m C(R_m)
\int d^d x\,
\overline G_\mu^a
\overline\partial^{\,2}\overline G^{\mu,a}
\\[0.5em]
&\quad
-\frac{1}{3}\sum_m C(R_m)
\int d^d x\,
\overline G_\mu^a
\overline\partial^\mu\overline\partial^\nu
\overline G_\nu^a
\\[0.5em]
&\quad
+\frac{g_s^2}{12}T_{\mathcal R}^{abcd}
\int d^d x\,
\overline G_\mu^a\overline G^{\mu,b}
\overline G_\nu^c\overline G^{\nu,d}
\\[0.5em]
&\quad
+\frac{g_s^2}{24}\sum_m C(R_m)\,f^{eac}f^{ebd}
\int d^d x\,
\overline G_\mu^a\overline G^{\mu,b}
\overline G_\nu^c\overline G^{\nu,d}
\\[0.5em]
&\quad
-\frac{\xi C_2(G)}{4}
\int d^d x\,
\left(
\overline R^{\,i}s_d\psi_{Ri}
+R^i s_d\overline\psi_{Ri}
\right)
\\[0.5em]
&\quad
+i\,\frac{5+\xi}{6}
\int d^d x\,
\overline\psi_i
\left[C_2(\mathcal R)\right]_{ij}
\overline{\slashed{\partial}}\,
\mathbb P_R\psi_j
\Bigg\}.
\end{aligned}
\end{equation}

Evanescent additions to $S_{\rm fct}^{(1)}$ do not affect one-loop symmetry restoration in the physical limit, but they can change the counter-terms required at two and higher loops.

The four-dimensional BRST variation of our finite counterterm
action agrees with Ref.~\cite{BeluscaMaito_2020}, after removing
the scalar and Yukawa sectors and identifying the group invariants
with those of the possibly reducible representation $\mathcal R$.
The actions differ by BRST-invariant terms: a contribution
proportional to
$\int d^4x\,\overline G_{\mu\nu}^a\overline G^{a,\mu\nu}$
and a sterile left-handed term
$$\int d^4x\,i\overline{\psi}_i
[C_2(\mathcal R)]_{ij}
\overline{\slashed{\partial}}\mathbb P_L\psi_j$$
included in that reference but absent in our choice.
The same agreement of BRST variations holds with
Ref.~\cite{Martin:1999cc}, after restricting to $SU(N)$
with right-handed fermions and translating its field normalization. However, arbitrary finite
BRST-invariant additions are explicitly allowed.

For intermediate breaking amplitudes, we find an evanescent
divergent $\psi\overline\psi c$ contribution absent from
Eq.~(45) of Ref.~\cite{Martin:1999cc}, and different colour
coefficients in its $GGGc$ amplitude, Eq.~(54).
The latter discrepancy was also noted in
Ref.~\cite{BeluscaMaito_2020}.
Despite these differences in the intermediate BRST-breaking
amplitudes, the equivalent finite counterterm action chosen
below agrees with the published final result after translating
the field normalization.

In general, we are free to add to our finite counter-terms any 
finite gauge invariant operator, since its BRST variation will vanish by construction and 
consequently it will not affect symmetry restoration. 

We choose the following equivalent form solely to facilitate
comparison with the literature and to fix the one-loop finite
counter-term convention used in our subsequent two-loop
comparison
:
\begin{equation}
\begin{aligned}
\label{eqn:fct-1-loop-inv}
\widetilde{S}_{\rm fct}^{(1)}
&=
\frac{g_s^2}{(4\pi)^2}
\Bigg\{
\frac{1}{4}\sum_m C(R_m)
\int d^d x\,
\overline G_\mu^a
\overline\partial^{\,2}\overline G^{\mu,a}
\\[0.5em]
&\quad
-\frac{5}{12}\sum_m C(R_m)
\int d^d x\,
\overline G_\mu^a
\overline\partial^\mu\overline\partial^\nu
\overline G_\nu^a
\\[0.5em]
&\quad
-\frac{g_s}{6}\sum_m C(R_m)\,f^{abc}
\int d^d x\,
\left(\overline\partial_\mu\overline G_\nu^a\right)
\overline G^{\mu,b}\overline G^{\nu,c}
\\[0.5em]
&\quad
+\frac{g_s^2}{12}T_{\mathcal R}^{abcd}
\int d^d x\,
\overline G_\mu^a\overline G^{\mu,b}
\overline G_\nu^c\overline G^{\nu,d}
\\[0.5em]
&\quad
-\frac{\xi C_2(G)}{4}
\int d^d x\,
\left(
\overline R^{\,i}s_d\psi_{Ri}
+R^i s_d\overline\psi_{Ri}
\right)
\\[0.5em]
&\quad
+i\,\frac{5+\xi}{6}
\int d^d x\,
\overline\psi_i
\left[C_2(\mathcal R)\right]_{ij}
\overline{\slashed{\partial}}\,
\mathbb P_R\psi_j
\Bigg\}.
\end{aligned}
\end{equation}
The difference in the physical
four-dimensional limit is
\begin{equation}
\begin{aligned}
\widetilde{S}_{\rm fct}^{(1)}-S_{\rm fct}^{(1)}
&=
-\frac{g_s^2}{(4\pi)^2}
\frac{\sum_m C(R_m)}{24}
\\
&\quad\times
\int d^4x\,
\overline G_{\mu\nu}^a\overline G^{a,\mu\nu},
\end{aligned}
\end{equation}
where
\begin{equation}
\overline G_{\mu\nu}^a
=
\overline\partial_\mu\overline G_\nu^a
-\overline\partial_\nu\overline G_\mu^a
+g_s f^{abc}\overline G_\mu^b\overline G_\nu^c.
\end{equation}
With the identification
$S_2(R)=C(\mathcal R)=\sum_m C(R_m)$ and the quadratic
Casimir understood as the matrix $C_2(\mathcal R)$,
the chosen solution agrees with the scalar- and Yukawa-free
result of Ref.~\cite{BeluscaMaito_2020} in the physical limit,
up to the BRST-invariant kinetic term for the sterile
left-handed fermions included in that reference.
Their four-dimensional BRST variations therefore coincide
for general $\xi$.

\subsection{Pole consistency check at one loop}

We can further verify that the BRST breaking generated by the singular counter-terms obtained through UV renormalization satisfies the divergent part of the restoration condition in Eq.~\eqref{eqn:1-loop-res_div}. 

\begin{equation}
\begin{aligned}
&\left.
\mathcal S_d\left(S_0+S_{\rm sct}^{(1)}\right)
\right|_{\text{1-loop}}
=&
-\frac{g_s^2}{(4\pi)^2}\frac{1}{2\epsilon}
\Bigg[
\xi C_2(G)\widehat{\Delta}
+\frac{2}{3}\sum_m C(R_m)
\int d^d x\,
\overline\partial_\mu c^a\,
\widehat\partial^{\,2}\overline G^{\mu,a}
\\[0.5em]
&&
+\frac{2g_s}{3}\sum_m C(R_m)\,f^{abc}
\int d^d x\,
c^a\overline G_\mu^c\,
\widehat\partial^{\,2}\overline G^{\mu,b}
\Bigg].
\end{aligned}
\end{equation}

This contribution exactly cancels the divergent breaking
$\left(\widehat{\Delta}\cdot\Gamma^{(1)}_{\rm div}\right)$
listed above, thereby providing a nontrivial consistency check of the singular counter-terms and of the one-loop restoration condition.
\par
The term proportional to $\widehat{\Delta}$ originates from the source-dependent singular counter-terms involving $R^i$ and $\overline{R}^{\,i}$. Through the nonlinear structure of the Slavnov--Taylor functional, these source counter-terms combine with the evanescent part of the tree-level fermionic kinetic term and reproduce the same local breaking operator $\widehat{\Delta}$.
\par
The remaining two terms are associated with the evanescent contribution proportional to
\begin{equation}
\int d^dx
\overline{G}_\mu^a
\widehat{\partial}^{\,2}
\overline{G}^{\mu,a}
\end{equation}
appearing in the singular part of the gauge-boson two-point function. Its Slavnov--Taylor variation generates both the term linear in the gauge field,
$\overline{\partial}_\mu c^a\,\widehat{\partial}^{\,2}\overline{G}^{\mu,a}$,
and the corresponding non-Abelian term proportional to
$g_sf^{abc}c^a\overline{G}_\mu^c\widehat{\partial}^{\,2}\overline{G}^{\mu,b}$.
Together, these contributions account for the complete divergent breaking generated by the $\widehat{\Delta}$ insertion.

\section{Two-loop analysis} 
\label{sec:twoloopsetup}
We now extend the previous analysis to two loops. 
Throughout the two-loop analysis\footnote{including the one-loop counter-terms entering the two-loop calculation. In particular, we are using eq.~\eqref{eqn:fct-1-loop-inv}, evaluated at $\xi=1$}, we restrict ourselves to the Feynman gauge,
$\xi=1$. 

As at one loop, UV renormalization requires the introduction of singular counter-terms $S_{\rm sct}^{(2)}$, while the restoration of the regularization-induced BRST breaking may additionally require finite symmetry-restoring counter-terms $S_{\rm fct}^{(2)}$. 

Including all contributions through two-loop order, the
renormalized effective action is
\begin{equation}
\begin{aligned}
\left.\Gamma_{\rm ren}\right|_{\text{through two loops}}
&=
S_0+\Gamma_{\rm}^{(1)}+S_{\rm ct}^{(1)}
+\Gamma^{(2)}
+S_{\rm ct}^{(2)}.
\end{aligned}
\end{equation}

Here, the contribution of total two-loop perturbative order to the effective action, $\Gamma^{(2)}$, includes genuine two-loop diagrams constructed from tree-level vertices, as well as one-loop diagrams
with one insertion of $S_{\rm ct}^{(1)}=S_{\rm sct}^{(1)}+S_{\rm fct}^{(1)}$. We denote, in general, $S_{\rm ct}^{(n)}=S_{\rm sct}^{(n)}+S_{\rm fct}^{(n)}$.

We seek the singular and finite two-loop counter-terms,
$S_{\rm sct}^{(2)}$ and $S_{\rm fct}^{(2)}$, such that
\begin{equation}
\lim_{d\to4}
\left\{
\mathcal S_d
\left(
S_0+\Gamma^{(1)}+S_{\rm sct}^{(1)}+S_{\rm fct}^{(1)}
+\Gamma^{(2)}+S_{\rm sct}^{(2)}+S_{\rm fct}^{(2)}
\right)
\right\}_{\rm 2-loop}
=0.
\end{equation}

Using manipulations analogous to those in
Eq.~\eqref{eq:SdManipulationsOneLoop}, and retaining only
terms of two-loop order, we obtain
\begin{equation}
\begin{aligned}
&
\left[
\mathcal S_d
\left(
S_0+\Gamma^{(1)}+S_{\rm ct}^{(1)}
+\Gamma^{(2)}+S_{\rm ct}^{(2)}
\right)
\right]_{\text{2-loop}}
\\[0.5em]
&\quad=
\left[
\mathcal S_d\left(S_0+\Gamma^{(1)}+\Gamma^{(2)}+S_{\rm ct}^{(1)}\right)
\right]_{\text{2-loop}}
+
\left[
\mathcal S_d
\left(S_0+S_{\rm ct}^{(2)}\right)
\right]_{\text{2-loop}}
\end{aligned}
\end{equation}

The nonlinear Slavnov--Taylor functional gives
\begin{equation}
\begin{aligned}
&
\left[
\mathcal S_d\left(
S_0+\Gamma^{(1)}+S_{\rm ct}^{(1)}+\Gamma^{(2)}
\right)
\right]_{\text{2-loop}}
\\
&=
\left[
\mathcal S_d\left(S_0+\Gamma^{(2)}\right)
\right]_{\text{2-loop}}
+
\frac{\delta\left(\Gamma^{(1)}+S_{\rm ct}^{(1)}\right)}
     {\delta X_i}
\circ
\frac{\delta\left(\Gamma^{(1)}+S_{\rm ct}^{(1)}\right)}
     {\delta\phi_i}.
\end{aligned}
\end{equation}
The QAP applies to this complete expression. We now express the QAP using the subgraph subtraction
prescription described above. The following relation is
understood for the local UV poles and physical finite
breaking contributions retained by this prescription.

\begin{equation}
\label{eq:QAP2loop}
\begin{aligned}
\left[
\mathcal S_d\left(S_0+\Gamma^{(1)}+\Gamma^{(2)}+S_{\rm ct}^{(1)}\right)
\right]_{\text{2-loop}}
&=
\left(\Delta\cdot\Gamma \right)_{\rm 2-loop}+\left[\mathcal S_d\left(S_{\rm ct}^{(1)}\right)\right]_{\text{2-loop}}.
\end{aligned}
\end{equation}
Although the last term on the right-hand side of this equation
is not written in the form $\Delta\cdot\Gamma$, its presence
follows from the inclusion of one-loop counter-terms and the
nonlinear nature of the Slavnov--Taylor functional.

As for the insertion operator $\Delta$, it contains the evanescent
BRST-breaking contributions from lower orders,
\begin{equation}
\Delta=\widehat{\Delta}+\widehat{\Delta}^{(1)},
\end{equation}
where $\widehat{\Delta}$ is the tree-level evanescent
BRST-breaking term obtained from $\mathcal S_d(S_0)$,
and $\widehat{\Delta}^{(1)}$ is the one-loop evanescent
BRST-breaking term obtained from
\begin{equation}
\widehat{\Delta}^{(1)}
=
\left[
\mathcal S_d
\left(
S_0+\Gamma^{(1)}+S_{\rm ct}^{(1)}
\right)
\right]_{\text{1-loop}}^{\rm evan,local}.
\end{equation}

The last term in Eq.~\eqref{eq:QAP2loop} is the contribution quadratic
in the one-loop counterterm action,
\begin{equation}
\left[
\mathcal S_d\left(S_{\rm ct}^{(1)}\right)
\right]_{\text{2-loop}}
=
\frac{\delta S_{\rm ct}^{(1)}}{\delta X_i}
\circ
\frac{\delta S_{\rm ct}^{(1)}}{\delta\phi_i}.
\end{equation}
It is only the counterterm--counterterm part of the quadratic
Slavnov--Taylor contribution. The remaining terms involve
$\Gamma^{(1)}$ and are included in the QAP relation.

The restoration procedure, at the previous perturbative level \eqref{eqn:BRST_inv_eff_action}, makes sure that the four-dimensional projection Eq.~\eqref{eqn:1-loop-res_fin}
is satisfied. The same holds for the divergent counterpart of this condition Eq.~\eqref{eqn:1-loop-res_div}. However, an evanescent remainder survives and
determines $\widehat{\Delta}^{(1)}$.
Using Eq.~\eqref{eq:SdManipulationsOneLoop}, we get
\begin{equation}
\widehat{\Delta}^{(1)}
=
\left[
\widehat{\Delta}\cdot\Gamma^{(1)}
\right]^{\rm evan,local}_{\rm 1-loop}+
\left[
\mathcal S_d\left(S_0+S_{\rm ct}^{(1)}\right)
\right]_{\text{1-loop}}^{\rm evan}.
\end{equation}

The finite one-loop source counter-terms produce the
evanescent variation
\begin{equation}
\begin{aligned}
\label{eqn:Delta1hat_ferm}
\left[
\mathcal S_d\!\left(S_0+S_{\rm fct}^{(1)}\right)
\right]_{\rm 1-loop}^{\rm evan}
&=
-\frac{g_s^3}{(4\pi)^2}
\frac{C_2(G)}{4}T_{\mathcal R,ij}^a
\int d^d x\,c^a
\left(
\overline{\psi}_i
\overleftarrow{\widehat{\slashed{\partial}}}
\mathbb{P}_R\psi_j
+
\overline{\psi}_i\mathbb{P}_L
\overrightarrow{\widehat{\slashed{\partial}}}\psi_j
\right).
\end{aligned}
\end{equation}
This contribution originates from the interplay between the finite
one-loop counter-terms involving the BRST sources $R^i$ and
$\overline R^{\,i}$ and the evanescent component of the tree-level
fermionic kinetic term. It gives the fermionic component of
$\widehat{\Delta}^{(1)}$, whose momentum-space Feynman rule is
\begin{equation}
\raisebox{-0.5\height}{
  \includegraphics[
    width=0.22\linewidth
  ]{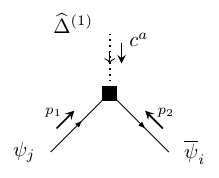}
}
\quad = -\frac{g_s^3}{(4\pi)^2}
T_{\mathcal R,ij}^a\,\frac{C_2(G)}{4}
\left(
\widehat{\slashed{p}}_1\mathbb{P}_L
+\widehat{\slashed{p}}_2\mathbb{P}_R
\right).
\end{equation}

Inspecting the one-loop computation, we see that the only contribution from $\left[
\widehat{\Delta}\cdot\Gamma_{\rm fin}^{(1)}
\right]^{\rm evan,local}$ comes from the finite evanescent part of
the one-loop gauge--ghost channel in Eq.~\eqref{eqn:Gc_viol_1_loop}.
Its local finite contribution contains the momentum structure
\begin{equation}
\overline p^\mu p^2
=
\overline p^\mu\overline p^{\,2}
+
\overline p^\mu\widehat p^{\,2}.
\end{equation}

Consequently, the complete effective local  insertion used in our
two-loop restoration equation is
\begin{equation}
\label{eqn:Delta1hat}
\boxed{
\widehat{\Delta}^{(1)}
=
\left[
\mathcal S_d\left(S_0+S_{\rm fct}^{(1)}\right)
\right]_{\rm 1\text{-}loop}^{\rm evan}
+
\left[
\widehat{\Delta}\cdot\Gamma_{\rm fin}^{(1)}
\right]_{Gc}^{\rm local,\,evan}.
}
\end{equation}

The origin of this effective insertion can be understood
directly from the subgraph subtraction. For a one-loop
subgraph containing $\widehat{\Delta}$, the subtraction
removes its local breaking contribution. Combining this
removed contribution with the Slavnov--Taylor variation
of the one-loop counterterms cancels the singular and
physical finite parts by the one-loop restoration
conditions. The remaining local evanescent contribution
is precisely $\widehat{\Delta}^{(1)}$. Throughout this
rearrangement, $\eta$ is kept independent of $\epsilon$
until the remaining loop integrations and pole extraction
have been completed.

We have to note here that the explicit insertion content depends on the subtraction
prescription. In Ref.~\cite{Belusca-Maito:2021lnk}, one-loop
counterterm diagrams and insertions of the BRST variation
of the one-loop counterterm action are included explicitly.
In particular, the variation of the finite counterterm
action provides a four-dimensional breaking insertion.
In our implementation, the singular counterterm contributions
and the contribution corresponding to such four-dimensional
breaking insertions are incorporated through subgraph
subtraction, which retains the finite terms generated by
evanescent numerator factors multiplying UV poles.
These contributions are therefore not added again separately.
Our effective insertion $\widehat{\Delta}^{(1)}$ contains
the local evanescent remainder left after one-loop symmetry
restoration. This organization is consistent with our
application of the quantum action principle above, understood
within the same subtraction prescription. Individual insertion
contributions must therefore be compared together with
their subtraction prescriptions.

Therefore, the two-loop restoration condition reads,
\begin{equation}
\label{eqn:rest-2-loop}
\lim_{d\rightarrow 4}\left(\widehat{\Delta}\cdot \Gamma^{(2)}+\widehat{\Delta}^{(1)}\cdot\Gamma^{(1)}+\mathcal{S}_{d}\left(S_{\rm ct}^{(1)} \right)_{\rm 2-loop}+\mathcal{S}_{d}\left(S_0+S_{\rm sct}^{(2)}+S_{\rm fct}^{(2)} \right)_{\rm 2-loop} \right) =0.
\end{equation}

The counter-term--counter-term contribution arises from the
quadratic part of the local action insertion. It is not
equal to the full quadratic term constructed from
$\Gamma_{\rm ren}^{(1)}$, which also contains products
involving one-loop diagrams.

\subsection{Two loop singular counter-terms}
The singular two-loop counter-terms cancel the UV divergences of the superficially divergent 1PI Green functions contained in $\Gamma^{(2)}$.

\begin{itemize}[leftmargin=0pt]
\item 
\textbf{Fermionic self-energy}:
\quad
\raisebox{-0.4\height}{
  \includegraphics[
    width=0.22\linewidth
  ]{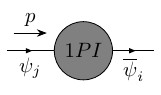}
}

\begin{equation}
\begin{aligned}
\left(S_{\text{sct,mom}}^{\psi\overline{\psi},(2)}\right)
&=
\frac{g_s^4}{(4\pi)^4}
\left\{
\frac{\sum_m C(R_m)}{9\epsilon}
\left[C_2(\mathcal R)\right]_{ij}
+C_2(G)\left[C_2(\mathcal R)\right]_{ij}
\left(
\frac{1}{\epsilon^2}-\frac{107}{36\epsilon}
\right)
\right.
\\[0.5em]
&\qquad\left.
+\left[C_2(\mathcal R)^2\right]_{ij}
\left(
\frac{1}{2\epsilon^2}+\frac{7}{12\epsilon}
\right)
\right\}
\overline{\slashed{p}}\mathbb{P}_R,
\end{aligned}
\end{equation}
\begin{equation}
\begin{aligned}
\left(S_{\text{sct}}^{\psi\overline{\psi},(2)}\right)&=
\frac{g_s^4}{(4\pi)^4}
\int d^d x\,\overline{\psi}_i
\left\{
\frac{\sum_m C(R_m)}{9\epsilon}
\left[C_2(\mathcal R)\right]_{ij}
+C_2(G)\left[C_2(\mathcal R)\right]_{ij}
\left(
\frac{1}{\epsilon^2}-\frac{107}{36\epsilon}
\right)
\right.
\\[0.5em]
&\qquad\left.
+\left[C_2(\mathcal R)^2\right]_{ij}
\left(
\frac{1}{2\epsilon^2}+\frac{7}{12\epsilon}
\right)
\right\}
i\,\overline{\slashed{\partial}}\,
\mathbb{P}_R\psi_j .
\end{aligned}
\end{equation}

\item 
\textbf{Gauge boson self-energy}:
\quad
\raisebox{-0.4\height}{
  \includegraphics[
    width=0.22\linewidth
  ]{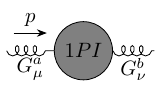}
}

\begin{equation}
\begin{aligned}
\left(S_{\text{sct,mom}}^{GG,(2)}\right)
&=
\frac{g_s^4}{(4\pi)^4}\delta^{ab}
\Bigg\{
C_2(G)^2
\left(
-\frac{25}{12\epsilon^2}+\frac{23}{8\epsilon}
\right)
\left(p^\mu p^\nu-p^2g^{\mu\nu}\right)
\\[0.5em]
&\quad
+2\sum_m C(R_m)C_2(R_m)
\Bigg[
\frac{1}{3\epsilon}
\left(
\overline{p}^{\,2}\overline{g}^{\mu\nu}
-\overline{p}^{\mu}\overline{p}^{\nu}
\right)
+
\left(
\frac{1}{12\epsilon^2}-\frac{17}{144\epsilon}
\right)
\widehat{p}^{\,2}\overline{g}^{\mu\nu}
\Bigg]
\\[0.5em]
&\quad
+2C_2(G)\sum_m C(R_m)
\Bigg[
\left(
\frac{5}{12\epsilon^2}-\frac{7}{8\epsilon}
\right)
\overline{p}^{\mu}\overline{p}^{\nu}
+
\left(
-\frac{1}{6\epsilon^2}+\frac{13}{36\epsilon}
\right)
\widehat{p}^{\mu}\widehat{p}^{\nu}
\\[0.5em]
&\qquad
+
\left(
\frac{5}{16\epsilon^2}-\frac{371}{576\epsilon}
\right)
\overline{p}^{\mu}\widehat{p}^{\nu}
+
\left(
\frac{5}{16\epsilon^2}-\frac{371}{576\epsilon}
\right)
\widehat{p}^{\mu}\overline{p}^{\nu}
\\[0.5em]
&\qquad
+
\left(
-\frac{5}{12\epsilon^2}+\frac{23}{36\epsilon}
\right)
\overline{p}^{\,2}\overline{g}^{\mu\nu}
+
\left(
-\frac{13}{36\epsilon^2}+\frac{103}{108\epsilon}
\right)
\overline{p}^{\,2}\widehat{g}^{\mu\nu}
\\[0.5em]
&\qquad
+
\left(
-\frac{7}{16\epsilon^2}+\frac{43}{576\epsilon}
\right)
\widehat{p}^{\,2}\overline{g}^{\mu\nu}
+
\left(
\frac{11}{72\epsilon^2}-\frac{317}{864\epsilon}
\right)
\widehat{p}^{\,2}\widehat{g}^{\mu\nu}
\Bigg]
\Bigg\},
\end{aligned}
\end{equation}
\begin{equation}
\begin{aligned}
\left(S_{\text{sct}}^{GG,(2)}\right)&=
\frac{g_s^4}{2(4\pi)^4}
\int d^d x\,G_\mu^a
\Bigg\{
C_2(G)^2
\left(
-\frac{25}{12\epsilon^2}+\frac{23}{8\epsilon}
\right)
\left(
g^{\mu\nu}\partial^2-\partial^\mu\partial^\nu
\right)
\\[0.5em]
&\quad
+2\sum_m C(R_m)C_2(R_m)
\Bigg[
\frac{1}{3\epsilon}
\left(
\overline{\partial}^{\mu}\overline{\partial}^{\nu}
-\overline{\partial}^{\,2}\overline{g}^{\mu\nu}
\right)
+
\left(
-\frac{1}{12\epsilon^2}+\frac{17}{144\epsilon}
\right)
\widehat{\partial}^{\,2}\overline{g}^{\mu\nu}
\Bigg]
\\[0.5em]
&\quad
+2C_2(G)\sum_m C(R_m)
\Bigg[
\left(
-\frac{5}{12\epsilon^2}+\frac{7}{8\epsilon}
\right)
\overline{\partial}^{\mu}\overline{\partial}^{\nu}
+
\left(
\frac{1}{6\epsilon^2}-\frac{13}{36\epsilon}
\right)
\widehat{\partial}^{\,\mu}\widehat{\partial}^{\,\nu}
\\[0.5em]
&\qquad
+
\left(
-\frac{5}{16\epsilon^2}+\frac{371}{576\epsilon}
\right)
\overline{\partial}^{\mu}\widehat{\partial}^{\,\nu}
+
\left(
-\frac{5}{16\epsilon^2}+\frac{371}{576\epsilon}
\right)
\widehat{\partial}^{\,\mu}\overline{\partial}^{\nu}
\\[0.5em]
&\qquad
+
\left(
\frac{5}{12\epsilon^2}-\frac{23}{36\epsilon}
\right)
\overline{\partial}^{\,2}\overline{g}^{\mu\nu}
+
\left(
\frac{13}{36\epsilon^2}-\frac{103}{108\epsilon}
\right)
\overline{\partial}^{\,2}\widehat{g}^{\mu\nu}
\\[0.5em]
&\qquad
+
\left(
\frac{7}{16\epsilon^2}-\frac{43}{576\epsilon}
\right)
\widehat{\partial}^{\,2}\overline{g}^{\mu\nu}
+
\left(
-\frac{11}{72\epsilon^2}+\frac{317}{864\epsilon}
\right)
\widehat{\partial}^{\,2}\widehat{g}^{\mu\nu}
\Bigg]
\Bigg\}
G_\nu^a .
\end{aligned}
\end{equation}

\item 
\textbf{Ghost self-energy}:
\quad
\raisebox{-0.4\height}{
  \includegraphics[
    width=0.22\linewidth
  ]{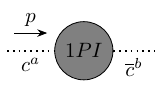}
}

\begin{equation}
\label{eqn:two_cc}
\begin{aligned}
\left(S_{\text{sct,mom}}^{c\overline{c},(2)}\right)
&=
\frac{g_s^4}{(4\pi)^4}\delta^{ab}
\Bigg\{
C_2(G)^2
\left(
-\frac{1}{\epsilon^2}+\frac{49}{48\epsilon}
\right)p^2 \\
&+
2C_2(G)\sum_m C(R_m)
\left(
\frac{1}{8\epsilon^2}-\frac{23}{72\epsilon}
\right)\overline{p}^{\,2}
\Bigg\},
\end{aligned}
\end{equation}
\begin{equation}
\begin{aligned}
\left(S_{\text{sct}}^{c\overline{c},(2)}\right)&=
\frac{g_s^4}{(4\pi)^4}
\int d^d x\,\overline{c}^{\,a}
\Bigg\{
C_2(G)^2
\left(
\frac{1}{\epsilon^2}-\frac{49}{48\epsilon}
\right)\partial^2
\\[0.5em]
&\quad
+
2C_2(G)\sum_m C(R_m)
\left(
-\frac{1}{8\epsilon^2}+\frac{23}{72\epsilon}
\right)\overline{\partial}^{\,2}
\Bigg\}
c^a .
\end{aligned}
\end{equation}

\item
\textbf{Ghost-BRST-source function}:
\quad
\raisebox{-0.4\height}{
  \includegraphics[
    width=0.22\linewidth
  ]{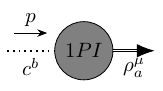}
}

\begin{equation}
\label{eqn:two_crho}
\begin{aligned}
\left(S_{\text{sct,mom}}^{\rho c,(2)}\right)
&=
\frac{i g_s^4}{(4\pi)^4}\delta^{ab}
\Bigg\{
C_2(G)^2
\left(
\frac{1}{\epsilon^2}-\frac{49}{48\epsilon}
\right)p^\mu \\
&+
2C_2(G)\sum_m C(R_m)
\left(
-\frac{1}{8\epsilon^2}+\frac{23}{72\epsilon}
\right)\overline{p}^{\mu}
\Bigg\},
\end{aligned}
\end{equation}
\begin{equation}
\begin{aligned}
\left(S_{\text{sct}}^{\rho c,(2)}\right)&=
\frac{g_s^4}{(4\pi)^4}
\int d^d x\,\rho_a^\mu
\Bigg\{
C_2(G)^2
\left(
-\frac{1}{\epsilon^2}+\frac{49}{48\epsilon}
\right)\partial_\mu
\\[0.5em]
&\quad
+
2C_2(G)\sum_m C(R_m)
\left(
\frac{1}{8\epsilon^2}-\frac{23}{72\epsilon}
\right)\overline{\partial}_\mu
\Bigg\}
c^a .
\end{aligned}
\end{equation}

\item 
\textbf{Fermion-gauge boson vertex}:
\quad
\raisebox{-0.3\height}{
  \includegraphics[
    width=0.3\linewidth
  ]{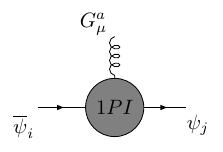}
}

\begin{equation}
\label{eqn:two_psipsiG}
\begin{aligned}
\left(S_{\text{sct,mom}}^{\psi\overline{\psi}G,(2)}\right)
&=
\frac{g_s^5}{(4\pi)^4}
\Bigg[
T_{\mathcal R}^a
\Bigg\{
C_2(G)^2
\left(
\frac{17}{8\epsilon^2}-\frac{67}{48\epsilon}
\right)\mathbf{1}
\\[0.5em]
&\quad
+C_2(G)\sum_m C(R_m)
\left(
-\frac{1}{4\epsilon^2}+\frac{8}{9\epsilon}
\right)\mathbf{1}
-\frac{\sum_m C(R_m)}{9\epsilon}C_2(\mathcal R)
\\[0.5em]
&\quad
+C_2(G)C_2(\mathcal R)
\left(
\frac{2}{\epsilon^2}-\frac{10}{9\epsilon}
\right)
+C_2(\mathcal R)^2
\left(
\frac{1}{2\epsilon^2}+\frac{17}{12\epsilon}
\right)
\Bigg\}
\Bigg]_{ij}
\overline{\gamma}^{\mu}\mathbb{P}_R,
\end{aligned}
\end{equation}
\begin{equation}
\begin{aligned}
\left(S_{\text{sct}}^{\psi\overline{\psi}G,(2)}\right)&=
\frac{g_s^5}{(4\pi)^4}
\int d^d x\,\overline{\psi}_i
\Bigg[
T_{\mathcal R}^a
\Bigg\{
C_2(G)^2
\left(
\frac{17}{8\epsilon^2}-\frac{67}{48\epsilon}
\right)\mathbf{1}
\\[0.5em]
&\quad
+C_2(G)\sum_m C(R_m)
\left(
-\frac{1}{4\epsilon^2}+\frac{8}{9\epsilon}
\right)\mathbf{1}
-\frac{\sum_m C(R_m)}{9\epsilon}C_2(\mathcal R)
\\[0.5em]
&\quad
+C_2(G)C_2(\mathcal R)
\left(
\frac{2}{\epsilon^2}-\frac{10}{9\epsilon}
\right)
+C_2(\mathcal R)^2
\left(
\frac{1}{2\epsilon^2}+\frac{17}{12\epsilon}
\right)
\Bigg\}
\Bigg]_{ij}
\overline{\gamma}^{\mu}\mathbb{P}_R\psi_j G_\mu^a .
\end{aligned}
\end{equation}

\item 
\textbf{Ghost-gauge boson vertex}:
\quad
\raisebox{-0.3\height}{
  \includegraphics[
    width=0.3\linewidth
  ]{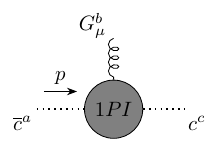}
}

\begin{equation}
\begin{aligned}
\left(S_{\mathrm{sct,mom}}^{\overline{c}cG,(2)}\right)
&=
\frac{i g_s^5}{(4\pi)^4}f^{abc}
\Bigg\{
C_2(G)^2
\left(
-\frac{5}{8\epsilon^2}+\frac{3}{8\epsilon}
\right)p^\mu
\\[0.5em]
&\qquad
+\frac{C_2(G)}{\epsilon}\sum_m C(R_m)
\left(
-\frac{\overline p^\mu}{24}
+\frac{\widehat p^\mu}{18}
\right)
\Bigg\},
\end{aligned}
\end{equation}
\begin{equation}
\begin{aligned}
\left(S_{\mathrm{sct}}^{\overline{c}cG,(2)}\right)&=
\frac{g_s^5}{(4\pi)^4}
\int d^d x\,f^{abc}
\Bigg\{
C_2(G)^2
\left(
\frac{5}{8\epsilon^2}-\frac{3}{8\epsilon}
\right)
\partial^\mu\overline c^{\,a}
\\[0.5em]
&\qquad
+\frac{C_2(G)}{\epsilon}\sum_m C(R_m)
\left(
\frac{\overline\partial^\mu\overline c^{\,a}}{24}
-\frac{\widehat\partial^\mu\overline c^{\,a}}{18}
\right)
\Bigg\}
G_\mu^b c^c .
\end{aligned}
\end{equation}

\item 
\textbf{Triple-gauge boson vertex}:
\quad
\raisebox{-0.3\height}{
  \includegraphics[
    width=0.3\linewidth
  ]{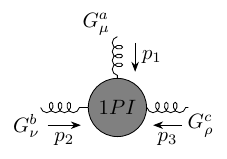}
}

\begin{equation}
\label{eqn:two_GGG}
\begin{aligned}
\left(S_{\text{sct,mom}}^{GGG,(2)}\right)
&=
\frac{i g_s^5}{(4\pi)^4}f^{abc}
\Bigg\{
C_2(G)^2
\left(
\frac{13}{8\epsilon^2}-\frac{71}{48\epsilon}
\right)
(p_1-p_2)^\rho g^{\mu\nu}
\\[0.5em]
&\quad
+2\sum_m C(R_m)
\left[
C_2(G)
\left(
-\frac{5}{8\epsilon^2}+\frac{67}{576\epsilon}
\right)
+\frac{C_2(R_m)}{6\epsilon}
\right]
\\[-0.2em]
&\qquad\times
(\overline{p}_1-\overline{p}_2)^\rho
\overline{g}^{\mu\nu}
\\[0.5em]
&\quad
+2C_2(G)\sum_m C(R_m)
\left(
-\frac{91}{384\epsilon^2}+\frac{2477}{4608\epsilon}
\right)
\\[-0.2em]
&\qquad\times
(\overline{p}_1-\overline{p}_2)^\rho
\widehat{g}^{\mu\nu}
\\[0.5em]
&\quad
+2C_2(G)\sum_m C(R_m)
\left(
-\frac{65}{384\epsilon^2}+\frac{619}{4608\epsilon}
\right)
\\[-0.2em]
&\qquad\times
(\widehat{p}_1-\widehat{p}_2)^\rho
\overline{g}^{\mu\nu}
\\[0.5em]
&\quad
+2C_2(G)\sum_m C(R_m)
\left(
\frac{7}{32\epsilon^2}-\frac{347}{1152\epsilon}
\right)
\\[-0.2em]
&\qquad\times
(\widehat{p}_1-\widehat{p}_2)^\rho
\widehat{g}^{\mu\nu}
+\text{cyclic permutations}
\Bigg\},
\end{aligned}
\end{equation}
\begin{equation}
\begin{aligned}
\left(S_{\text{sct}}^{GGG,(2)}\right)&=
\frac{g_s^5}{(4\pi)^4}
\int d^d x\,f^{abc}
\Bigg\{
C_2(G)^2
\left(
\frac{13}{8\epsilon^2}-\frac{71}{48\epsilon}
\right)
\\[-0.2em]
&\qquad\times
\left(\partial^\mu G_\nu^a\right)
G_\mu^b G^{c\nu}
\\[0.5em]
&\quad
+2\sum_m C(R_m)
\left[
C_2(G)
\left(
-\frac{5}{8\epsilon^2}+\frac{67}{576\epsilon}
\right)
+\frac{C_2(R_m)}{6\epsilon}
\right]
\\[-0.2em]
&\qquad\times
\left(\partial^\mu\overline{G}_\nu^a\right)
\overline{G}_\mu^b\overline{G}^{c\nu}
\\[0.5em]
&\quad
+2C_2(G)\sum_m C(R_m)
\left(
-\frac{91}{384\epsilon^2}+\frac{2477}{4608\epsilon}
\right)
\\[-0.2em]
&\qquad\times
\left(\partial^\mu\widehat{G}_\nu^a\right)
\overline{G}_\mu^b\widehat{G}^{c\nu}
\\[0.5em]
&\quad
+2C_2(G)\sum_m C(R_m)
\left(
-\frac{65}{384\epsilon^2}+\frac{619}{4608\epsilon}
\right)
\\[-0.2em]
&\qquad\times
\left(\partial^\mu\overline{G}_\nu^a\right)
\widehat{G}_\mu^b\overline{G}^{c\nu}
\\[0.5em]
&\quad
+2C_2(G)\sum_m C(R_m)
\left(
\frac{7}{32\epsilon^2}-\frac{347}{1152\epsilon}
\right)
\\[-0.2em]
&\qquad\times
\left(\partial^\mu\widehat{G}_\nu^a\right)
\widehat{G}_\mu^b\widehat{G}^{c\nu}
\Bigg\}.
\end{aligned}
\end{equation}

\item 
\textbf{Ghost-gluon-BRST-source vertex}:
\quad
\raisebox{-0.3\height}{
  \includegraphics[
    width=0.3\linewidth
  ]{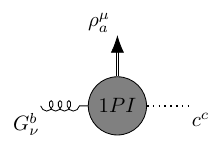}
}

\begin{equation}
\label{eqn:two_cGrho}
\begin{aligned}
\left(S_{\text{sct,mom}}^{\rho cG,(2)}\right)
&=
\frac{g_s^5}{(4\pi)^4}f^{abc}
\Bigg\{
C_2(G)^2
\left(
\frac{5}{8\epsilon^2}-\frac{3}{8\epsilon}
\right)g^{\mu\nu}
\\[0.5em]
&\qquad
+\frac{C_2(G)}{\epsilon}\sum_m C(R_m)
\left(
\frac{\overline{g}^{\mu\nu}}{24}
-\frac{\widehat{g}^{\mu\nu}}{18}
\right)
\Bigg\},
\end{aligned}
\end{equation}
\begin{equation}
\begin{aligned}
\left(S_{\text{sct}}^{\rho cG,(2)}\right)&=
\frac{g_s^5}{(4\pi)^4}
\int d^d x\,f^{abc}\,
\rho_a^\mu G^{b\nu}c^c
\Bigg\{
C_2(G)^2
\left(
\frac{5}{8\epsilon^2}-\frac{3}{8\epsilon}
\right)g_{\mu\nu}
\\[0.5em]
&\qquad
+\frac{C_2(G)}{\epsilon}\sum_m C(R_m)
\left(
\frac{\overline{g}_{\mu\nu}}{24}
-\frac{\widehat{g}_{\mu\nu}}{18}
\right)
\Bigg\}.
\end{aligned}
\end{equation}

\item 
\textbf{Ghost-ghost-BRST-source vertex}:
\quad
\raisebox{-0.3\height}{
  \includegraphics[
    width=0.3\linewidth
  ]{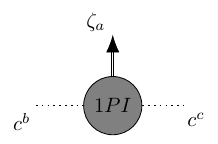}
}

\begin{equation}
\label{eqn:two_cczeta}
\begin{aligned}
\left(S_{\text{sct,mom}}^{\zeta cc,(2)}\right)
&=
\frac{g_s^5}{(4\pi)^4}f^{abc}
\Bigg\{
C_2(G)^2
\left(
-\frac{5}{8\epsilon^2}+\frac{3}{8\epsilon}
\right)
\\[0.5em]
&\qquad
+\frac{C_2(G)}{18\epsilon}\sum_m C(R_m)
\Bigg\},
\end{aligned}
\end{equation}
\begin{equation}
\begin{aligned}
\left(S_{\text{sct}}^{\zeta cc,(2)}\right)&=
\frac{g_s^5}{2(4\pi)^4}
\int d^d x\,f^{abc}\,
\zeta_a\,c^b c^c
\Bigg\{
C_2(G)^2
\left(
-\frac{5}{8\epsilon^2}+\frac{3}{8\epsilon}
\right)
\\[0.5em]
&\qquad
+\frac{C_2(G)}{18\epsilon}\sum_m C(R_m)
\Bigg\}.
\end{aligned}
\end{equation}
 
\item
\textbf{Fermion-ghost-BRST-source vertex}:
\quad
\raisebox{-0.3\height}{
  \includegraphics[
    width=0.3\linewidth
  ]{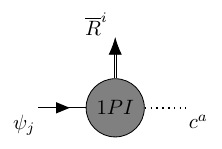}
}

\begin{equation}
\label{eqn:twopsicRbar}
\begin{aligned}
\left(S_{\text{sct,mom}}^{\overline{R}c\psi,(2)}\right)
&=
\frac{i g_s^5}{(4\pi)^4}
\Bigg[
T_{\mathcal R}^a
\Bigg\{
\frac{5C_2(G)^2}{8\epsilon^2}\mathbf{1}
-\frac{C_2(G)}{18\epsilon}
\sum_m C(R_m)\,\mathbf{1}
\\[0.5em]
&\qquad
+\frac{C_2(G)}{6\epsilon}C_2(\mathcal R)
\Bigg\}
\Bigg]_{ij}\mathbb{P}_R,
\end{aligned}
\end{equation}
\begin{equation}
\begin{aligned}
\left(S_{\text{sct}}^{\overline{R}c\psi,(2)}\right)&=
\frac{i g_s^5}{(4\pi)^4}
\int d^d x\,\overline{R}^{\,i}
\Bigg[
T_{\mathcal R}^a
\Bigg\{
\frac{5C_2(G)^2}{8\epsilon^2}\mathbf{1}
-\frac{C_2(G)}{18\epsilon}
\sum_m C(R_m)\,\mathbf{1}
\\[0.5em]
&\qquad
+\frac{C_2(G)}{6\epsilon}C_2(\mathcal R)
\Bigg\}
\Bigg]_{ij}
\mathbb{P}_R\psi_j\,c^a .
\end{aligned}
\end{equation}

\item 
\textbf{Anti-Fermion-ghost-BRST-source vertex}:
\quad
\raisebox{-0.3\height}{
  \includegraphics[
    width=0.3\linewidth
  ]{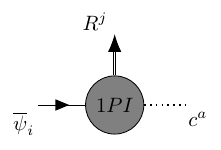}
}

\begin{equation}
\label{eqn:two_psicRbar}
\begin{aligned}
\left(S_{\text{sct,mom}}^{Rc\overline{\psi},(2)}\right)
&=
\frac{i g_s^5}{(4\pi)^4}
\Bigg[
T_{\mathcal R}^a
\Bigg\{
\frac{5C_2(G)^2}{8\epsilon^2}\mathbf{1}
-\frac{C_2(G)}{18\epsilon}
\sum_m C(R_m)\,\mathbf{1}
\\[0.5em]
&\qquad
+\frac{C_2(G)}{6\epsilon}C_2(\mathcal R)
\Bigg\}
\Bigg]_{ij}\mathbb{P}_L,
\end{aligned}
\end{equation}
\begin{equation}
\begin{aligned}
\left(S_{\text{sct}}^{Rc\overline{\psi},(2)}\right)&=
\frac{i g_s^5}{(4\pi)^4}
\int d^d x\,\overline{\psi}_i
\Bigg[
T_{\mathcal R}^a
\Bigg\{
\frac{5C_2(G)^2}{8\epsilon^2}\mathbf{1}
-\frac{C_2(G)}{18\epsilon}
\sum_m C(R_m)\,\mathbf{1}
\\[0.5em]
&\qquad
+\frac{C_2(G)}{6\epsilon}C_2(\mathcal R)
\Bigg\}
\Bigg]_{ij}
\mathbb{P}_L R^j\,c^a .
\end{aligned}
\end{equation}

\item 
\textbf{Four-gluon vertex}:
\quad
\raisebox{-0.5\height}{
  \includegraphics[
    width=0.22\linewidth
  ]{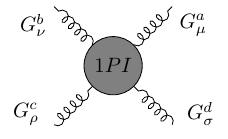}
}

The representation sums and ordered traces follow the group-theory
conventions introduced above.

\begin{equation}
\begin{aligned}
S^{GGGG,(2)}_{\mathrm{sct,mom}}
&=
\frac{g_s^6}{(4\pi)^4}
\Bigg\{
C_2(G)^2\mathcal F^{abcd}
\left(
\frac{1}{3\epsilon^2}-\frac{1}{6\epsilon}
\right)
g^{\mu\nu}g^{\rho\sigma}
\\[0.5em]
&\quad
+\overline g^{\mu\nu}\overline g^{\rho\sigma}
\Bigg[
\frac{\mathcal Q_{\mathcal R}^{ab|cd}}{6\epsilon}
+\frac{2}{3\epsilon}
\left(
\mathcal Q_{\mathcal R}^{ac|bd}
+\mathcal Q_{\mathcal R}^{ad|bc}
\right)
\\[0.3em]
&\qquad
+C_2(G)\mathcal C_{\mathcal R}^{abcd}
\left(
-\frac{5}{3\epsilon^2}-\frac{389}{288\epsilon}
\right)
\\[0.3em]
&\qquad
+C_2(G)\mathcal D_{\mathcal R}^{abcd}
\left(
\frac{10}{3\epsilon^2}+\frac{281}{144\epsilon}
\right)
\Bigg]
\\[0.5em]
&\quad
+\left(
\overline g^{\mu\nu}\widehat g^{\rho\sigma}
+\widehat g^{\mu\nu}\overline g^{\rho\sigma}
\right)
\Bigg[
\mathcal Q_{\mathcal R}^{ab|cd}
\left(
-\frac{11}{48\epsilon^2}+\frac{185}{576\epsilon}
\right)
\\[0.3em]
&\qquad
+\left(
\mathcal Q_{\mathcal R}^{ac|bd}
+\mathcal Q_{\mathcal R}^{ad|bc}
\right)
\left(
-\frac{11}{48\epsilon^2}+\frac{281}{576\epsilon}
\right)
\\[0.3em]
&\qquad
+C_2(G)\mathcal C_{\mathcal R}^{abcd}
\left(
-\frac{19}{72\epsilon^2}+\frac{389}{1728\epsilon}
\right)
\\[0.3em]
&\qquad
+C_2(G)\mathcal D_{\mathcal R}^{abcd}
\left(
\frac{119}{288\epsilon^2}-\frac{1001}{3456\epsilon}
\right)
\Bigg]
\\[0.5em]
&\quad
+\widehat g^{\mu\nu}\widehat g^{\rho\sigma}
\Bigg[
\left(
\mathcal Q_{\mathcal R}^{ab|cd}
+\mathcal Q_{\mathcal R}^{ac|bd}
+\mathcal Q_{\mathcal R}^{ad|bc}
\right)
\left(
-\frac{11}{36\epsilon^2}-\frac{139}{432\epsilon}
\right)
\\[0.3em]
&\qquad
+C_2(G)\mathcal C_{\mathcal R}^{abcd}
\left(
\frac{73}{144\epsilon^2}-\frac{919}{1728\epsilon}
\right)
\\[0.3em]
&\qquad
+C_2(G)\mathcal D_{\mathcal R}^{abcd}
\left(
-\frac{7}{6\epsilon^2}+\frac{65}{72\epsilon}
\right)
\Bigg]
\\[0.5em]
&\quad
+\begin{pmatrix}
\nu\leftrightarrow\rho\\
b\leftrightarrow c
\end{pmatrix}
+\begin{pmatrix}
\nu\leftrightarrow\sigma\\
b\leftrightarrow d
\end{pmatrix}
\Bigg\}.
\end{aligned}
\end{equation}

\begin{equation}
\begin{aligned}
S^{GGGG,(2)}_{\mathrm{sct}}
&=
\frac{g_s^6}{8(4\pi)^4}
\int d^d x\,
G_\mu^a G_\nu^b G_\rho^c G_\sigma^d
\\[0.3em]
&\quad\times
\Bigg\{
C_2(G)^2\mathcal F^{abcd}
\left(
\frac{1}{3\epsilon^2}-\frac{1}{6\epsilon}
\right)
g^{\mu\nu}g^{\rho\sigma}
\\[0.5em]
&\quad
+\overline g^{\mu\nu}\overline g^{\rho\sigma}
\Bigg[
\frac{\mathcal Q_{\mathcal R}^{ab|cd}}{6\epsilon}
+\frac{2}{3\epsilon}
\left(
\mathcal Q_{\mathcal R}^{ac|bd}
+\mathcal Q_{\mathcal R}^{ad|bc}
\right)
\\[0.3em]
&\qquad
+C_2(G)\mathcal C_{\mathcal R}^{abcd}
\left(
-\frac{5}{3\epsilon^2}-\frac{389}{288\epsilon}
\right)
\\[0.3em]
&\qquad
+C_2(G)\mathcal D_{\mathcal R}^{abcd}
\left(
\frac{10}{3\epsilon^2}+\frac{281}{144\epsilon}
\right)
\Bigg]
\\[0.5em]
&\quad
+\left(
\overline g^{\mu\nu}\widehat g^{\rho\sigma}
+\widehat g^{\mu\nu}\overline g^{\rho\sigma}
\right)
\\[0.3em]
&\qquad\times
\Bigg[
\mathcal Q_{\mathcal R}^{ab|cd}
\left(
-\frac{11}{48\epsilon^2}+\frac{185}{576\epsilon}
\right)
\\[0.3em]
&\qquad
+\left(
\mathcal Q_{\mathcal R}^{ac|bd}
+\mathcal Q_{\mathcal R}^{ad|bc}
\right)
\left(
-\frac{11}{48\epsilon^2}+\frac{281}{576\epsilon}
\right)
\\[0.3em]
&\qquad
+C_2(G)\mathcal C_{\mathcal R}^{abcd}
\left(
-\frac{19}{72\epsilon^2}+\frac{389}{1728\epsilon}
\right)
\\[0.3em]
&\qquad
+C_2(G)\mathcal D_{\mathcal R}^{abcd}
\left(
\frac{119}{288\epsilon^2}-\frac{1001}{3456\epsilon}
\right)
\Bigg]
\\[0.5em]
&\quad
+\widehat g^{\mu\nu}\widehat g^{\rho\sigma}
\\[0.3em]
&\qquad\times
\Bigg[
\left(
\mathcal Q_{\mathcal R}^{ab|cd}
+\mathcal Q_{\mathcal R}^{ac|bd}
+\mathcal Q_{\mathcal R}^{ad|bc}
\right)
\left(
-\frac{11}{36\epsilon^2}-\frac{139}{432\epsilon}
\right)
\\[0.3em]
&\qquad
+C_2(G)\mathcal C_{\mathcal R}^{abcd}
\left(
\frac{73}{144\epsilon^2}-\frac{919}{1728\epsilon}
\right)
\\[0.3em]
&\qquad
+C_2(G)\mathcal D_{\mathcal R}^{abcd}
\left(
-\frac{7}{6\epsilon^2}+\frac{65}{72\epsilon}
\right)
\Bigg]
\Bigg\}.
\end{aligned}
\end{equation}

where we have defined the following color tensors

\begin{equation}
\begin{aligned}
\mathcal C_X^{abcd}
&=
T_X^{abcd}+T_X^{adcb}
+T_X^{abdc}+T_X^{acdb},
\\[0.5em]
\mathcal D_X^{abcd}
&=
T_X^{adbc}+T_X^{acbd},
\\[0.5em]
\mathcal F^{abcd}
&=
\frac12
\left(
f^{ace}f^{bde}
+f^{ade}f^{bce}
\right),
\\[0.5em]
2C(X)\mathcal F^{abcd}
&=
T_X^{abcd}+T_X^{adcb}
+T_X^{abdc}+T_X^{acdb}
-2T_X^{adbc}-2T_X^{acbd}
\\[0.5em]
&=
\mathcal C_X^{abcd}-2\mathcal D_X^{abcd},
\qquad X=\mathcal R,A,
\\[0.5em]
C(\mathcal R)
&=\sum_m C(R_m),
\qquad C(A)=C_2(G),
\\[0.5em]
\mathcal Q_{\mathcal R}^{ab|cd}
&=
\frac13 f^{aex}f^{bfx}
\left(
\mathcal C_{\mathcal R}^{cdef}
+\mathcal D_{\mathcal R}^{cdef}
\right)
-\frac{C_2(G)}6
\left(
\mathcal C_{\mathcal R}^{abcd}
+\mathcal D_{\mathcal R}^{abcd}
\right)
\\[0.5em]
&\quad
+\frac{C(\mathcal R)}{18}
\left(
\mathcal C_A^{abcd}
+\mathcal D_A^{abcd}
\right)
-\frac{C_2(G)C(\mathcal R)}9\mathcal F^{abcd}.
\end{aligned}
\end{equation}
Equivalently,
\begin{equation}
\begin{aligned}
\mathcal Q_{\mathcal R}^{ab|cd}
&=\sum_m\mathcal Q_{R_m}^{ab|cd}
\\[0.5em]
&=\sum_m\Bigg\{
\frac12\sum_e
\left(
T_{R_m}^{abecde}+T_{R_m}^{baecde}
+T_{R_m}^{abedce}+T_{R_m}^{baedce}
\right)
+\frac{C_2(G)-2C_2(R_m)}4
\mathcal C_{R_m}^{abcd}
\Bigg\}.
\end{aligned}
\end{equation}
These trace tensors form a convenient spanning set for the
colour structures appearing in our results. They need not be
independent, additional relations can hold for particular
gauge groups or representations.

\end{itemize}

Finally, all results presented above are found to agree with the known two-loop results of Ref.~\cite{Kuhler:2025znv} upon taking the corresponding $SU(2)$ limit and matching the conventions and parameter choices used therein.

\subsection{BRST violating terms at two loops}

Below, we list the finite four-dimensional contributions associated with all superficially divergent 1PI Green functions whose local breaking can contribute up to canonical dimension four and ghost number one.

For convenience, we define
\begin{equation}
\widetilde{\widehat{\Delta}}\cdot\Gamma^{(2)}
=
\widehat{\Delta}\cdot \Gamma^{(2)}
+
\widehat{\Delta}^{(1)}\cdot\Gamma^{(1)},
\end{equation}
and present the corresponding results in both momentum and position space.

Following Sec.~\ref{sec:twoloopsetup}, we generate diagrams
for all candidate 1PI channels listed in
Table~\ref{tab:brst-breaking-channels}, including those
first appearing at two-loop order. The results below
include the following diagrammatic contributions
to the two-loop BRST breaking:
\begin{itemize}
    \item Two-loop diagrams with one $\widehat{\Delta}$ insertion
    and otherwise tree-level vertices.
    \item One-loop diagrams with one $S_{\rm fct}^{(1)}$ insertion,
    one $\widehat{\Delta}$ insertion, and otherwise tree-level vertices.
    \item One-loop diagrams with one $\widehat{\Delta}^{(1)}$
    insertion and otherwise tree-level vertices.
\end{itemize}

The position and momentum-space expressions in this list are understood after
physical four-dimensional projection. We retain $d^dx$ as notation
for the local representatives before this projection.

\begin{itemize}[leftmargin=0pt]

\item
\textbf{Gauge-boson-ghost function}:
\quad
\raisebox{-0.4\height}{
  \includegraphics[
    width=0.22\linewidth
  ]{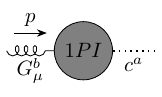}
}

\begin{equation}
\begin{aligned}
\lim_{d\rightarrow4}
\left(
\widetilde{\widehat{\Delta}}
\cdot\Gamma_{\mathrm{fin}}^{(2)}
\right)_{Gc,\text{mom}}
&=
-\frac{i g_s^4}{(4\pi)^4}\,
\delta^{ab}\,2\sum_m C(R_m)
\left(
\frac{7}{108}C_2(G)
-\frac{11}{48}C_2(R_m)
\right)
\\[0.5em]
&\qquad\times
\overline{p}^{\mu}\overline{p}^{\,2},
\end{aligned}
\end{equation}
\begin{equation}
\begin{aligned}
\lim_{d\rightarrow4}
\left(
\widetilde{\widehat{\Delta}}
\cdot\Gamma_{\mathrm{fin}}^{(2)}
\right)_{Gc}&=
-\frac{g_s^4}{(4\pi)^4}\,
2\sum_m C(R_m)
\left(
\frac{7}{108}C_2(G)
-\frac{11}{48}C_2(R_m)
\right)
\\[0.5em]
&\qquad\times
\int d^d x\,c^a\,
\overline{\partial}_{\mu}
\overline{\partial}^{\,2}
\overline{G}^{\mu,a}.
\end{aligned}
\end{equation}

\item 
\textbf{Double-gauge-boson-ghost vertex}:
\quad
\raisebox{-0.3\height}{
  \includegraphics[
    width=0.3\linewidth
  ]{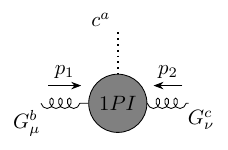}
}

\begin{equation}
\begin{aligned}
\lim_{d\rightarrow4}
\left(
\widetilde{\widehat{\Delta}}
\cdot\Gamma_{\mathrm{fin}}^{(2)}
\right)_{GGc,\text{mom}}
&=
-\frac{g_s^5}{(4\pi)^4}\,
2f^{abc}\sum_m C(R_m)
\Bigg[
\left(
\frac{271}{2304}C_2(G)
-\frac{5}{72}C_2(R_m)
\right)
\\[0.5em]
&\qquad\times
\left(
\overline{p}_1^{\,2}-\overline{p}_2^{\,2}
\right)\overline{g}^{\mu\nu}
\\[0.5em]
&\quad
+\left(
-\frac{365}{6912}C_2(G)
-\frac{23}{144}C_2(R_m)
\right)
\\[0.5em]
&\qquad\times
\left(
\overline{p}_1^\mu\overline{p}_1^\nu
-\overline{p}_2^\mu\overline{p}_2^\nu
\right)
\Bigg],
\end{aligned}
\end{equation}
\begin{equation}
\begin{aligned}
\lim_{d\rightarrow4}
\left(
\widetilde{\widehat{\Delta}}
\cdot\Gamma_{\mathrm{fin}}^{(2)}
\right)_{GGc}&=
-\frac{g_s^5}{(4\pi)^4}\,
2f^{abc}\sum_m C(R_m)
\Bigg[
\left(
-\frac{271}{2304}C_2(G)
+\frac{5}{72}C_2(R_m)
\right)
\\[0.5em]
&\qquad\times
\int d^d x\,c^a\,
\overline{G}_\mu^c\,
\overline{\partial}^{\,2}
\overline{G}^{\mu,b}
\\[0.5em]
&\quad
+\left(
\frac{365}{6912}C_2(G)
+\frac{23}{144}C_2(R_m)
\right)
\\[0.5em]
&\qquad\times
\int d^d x\,c^a\,
\overline{G}_\nu^c\,
\overline{\partial}^{\,\mu}
\overline{\partial}^{\,\nu}
\overline{G}_\mu^b
\Bigg].
\end{aligned}
\end{equation}

\item 
\textbf{Triple-gauge-boson-ghost vertex}:
\quad
\raisebox{-0.4\height}{
  \includegraphics[
    width=0.3\linewidth
  ]{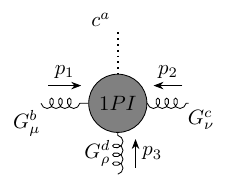}
}

\begin{equation}
\begin{aligned}
\lim_{d\rightarrow4}
\left(
\widetilde{\widehat{\Delta}}
\cdot\Gamma_{\rm fin}^{(2)}
\right)_{GGGc,\text{mom}}
&=
-\frac{i g_s^6}{(4\pi)^4}
\left(\overline p_1+\overline p_2+\overline p_3\right)_\sigma
\overline g^{\mu\nu}\overline g^{\rho\sigma}
\\[0.5em]
&\quad\times
\sum_m\Bigg\{
\frac{5}{72}
\left[
\mathcal Q_{R_m}^{ad|bc}
-2\mathcal Q_{R_m}^{ac|bd}
-2\mathcal Q_{R_m}^{ab|cd}
\right]
\\[0.5em]
&\qquad
+\left(
-\frac{235}{864}C_2(G)
-\frac{17}{72}C_2(R_m)
\right)\mathcal C_{R_m}^{adbc}
\\[0.5em]
&\qquad
+\left(
\frac{10}{27}C_2(G)
-\frac{5}{18}C_2(R_m)
\right)\mathcal D_{R_m}^{adbc}
\Bigg\}
\\[0.5em]
&\quad
+\begin{pmatrix}
\nu\leftrightarrow\rho\\
c\leftrightarrow d
\end{pmatrix}
+\begin{pmatrix}
\mu\leftrightarrow\rho\\
b\leftrightarrow d
\end{pmatrix},
\end{aligned}
\end{equation}
\begin{equation}
\begin{aligned}
\lim_{d\rightarrow4}
\left(
\widetilde{\widehat{\Delta}}
\cdot\Gamma_{\rm fin}^{(2)}
\right)_{GGGc}&=
-\frac{g_s^6}{(4\pi)^4}
\sum_m\Bigg\{
-\frac{5}{144}
\left[
\mathcal Q_{R_m}^{ad|bc}
-2\mathcal Q_{R_m}^{ac|bd}
-2\mathcal Q_{R_m}^{ab|cd}
\right]
\\[0.5em]
&\qquad
+\left(
\frac{235}{1728}C_2(G)
+\frac{17}{144}C_2(R_m)
\right)\mathcal C_{R_m}^{adbc}
\\[0.5em]
&\qquad
+\left(
-\frac{5}{27}C_2(G)
+\frac{5}{36}C_2(R_m)
\right)\mathcal D_{R_m}^{adbc}
\Bigg\}
\\[0.5em]
&\quad\times
\int d^d x\,c^a\,\overline\partial^\nu
\left(
\overline G_\mu^b
\overline G^{\mu,c}
\overline G_\nu^d
\right).
\end{aligned}
\end{equation}
Here $\mathcal C_{R_m}$, $\mathcal D_{R_m}$, and
$\mathcal Q_{R_m}$ are the trace tensors defined previously,
evaluated in each irreducible block $R_m$ of
$\mathcal R=\bigoplus_m R_m$.

\item
\textbf{Quadruple-gauge-boson-ghost vertex}:
\quad
\raisebox{-0.42\height}{
  \includegraphics[
    width=0.22\linewidth
  ]{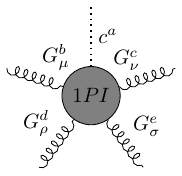}
}

\begin{equation}
\begin{aligned}
    \lim_{d\rightarrow 4}\left(\widetilde{\widehat{\Delta}}\cdot\Gamma_{\rm fin}^{(2)}\right)_{GGGGc} = 0,
\end{aligned}
\end{equation}
This vanishing is obtained after summing the contributing
insertion classes and imposing anomaly cancellation on the
total fermion representation. The role of this condition in
the antisymmetrized colour traces can be seen from
\begin{equation}
\begin{aligned}
&\operatorname{Tr}_{\mathcal R}
\left(
T_{\mathcal R}^a
T_{\mathcal R}^{[b}
T_{\mathcal R}^{c}
T_{\mathcal R}^{d}
T_{\mathcal R}^{e]}
\right)
\\
&\qquad =
-\frac{1}{12}
\left(
f^{bcx}f^{dey}
-f^{bdx}f^{cey}
+f^{bex}f^{cdy}
\right)
d_{\mathcal R}^{axy},
\end{aligned}
\end{equation}
where the brackets denote antisymmetrization over $b,c,d,e$
with weight $1/4!$, and $d_{\mathcal R}^{axy}$ is the
symmetrized cubic trace defined in Appendix~A.
Since
\begin{equation}
d_{\mathcal R}^{axy}
=
\sum_m d_{R_m}^{axy}
=
0,
\end{equation}
these trace contributions vanish. The remaining terms cancel
using the Lie-algebra and trace identities. 

\item 
\textbf{Fermion-antifermion-ghost vertex}:
\quad
\raisebox{-0.3\height}{
  \includegraphics[
    width=0.3\linewidth
  ]{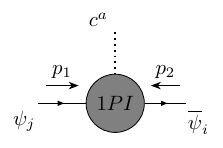}
}

\begin{equation}
\begin{aligned}
\lim_{d\rightarrow4}
\left(
\widetilde{\widehat{\Delta}}
\cdot\Gamma_{\mathrm{fin}}^{(2)}
\right)_{\psi\overline{\psi}c,\text{mom}}
&=
-\frac{i g_s^5}{(4\pi)^4}
\Bigg[
T_{\mathcal R}^{a}
\Bigg\{
\frac{4}{3}C_2(G)^2\mathbf{1}
-\frac{1121}{216}C_2(G)C_2(\mathcal R)
\\[0.5em]
&\qquad
+\frac{127}{36}C_2(\mathcal R)^2
-\frac{55}{216}C_2(G)
\sum_m C(R_m)\,\mathbf{1}
\\[0.5em]
&\qquad
+\frac{41}{108}
\left(\sum_m C(R_m)\right)C_2(\mathcal R)
\Bigg\}
\Bigg]_{ij}
\\[0.5em]
&\quad\times
\left(
\overline{\slashed{p}}_1
+\overline{\slashed{p}}_2
\right)\mathbb{P}_R,
\end{aligned}
\end{equation}
\begin{equation}
\begin{aligned}
\lim_{d\rightarrow4}
\left(
\widetilde{\widehat{\Delta}}
\cdot\Gamma_{\mathrm{fin}}^{(2)}
\right)_{\psi\overline{\psi}c}&=
\frac{g_s^5}{(4\pi)^4}
\Bigg[
T_{\mathcal R}^{a}
\Bigg\{
\frac{4}{3}C_2(G)^2\mathbf{1}
-\frac{1121}{216}C_2(G)C_2(\mathcal R)
\\[0.5em]
&\qquad
+\frac{127}{36}C_2(\mathcal R)^2
-\frac{55}{216}C_2(G)
\sum_m C(R_m)\,\mathbf{1}
\\[0.5em]
&\qquad
+\frac{41}{108}
\left(\sum_m C(R_m)\right)C_2(\mathcal R)
\Bigg\}
\Bigg]_{ij}
\\[0.5em]
&\quad\times
\int d^d x\,c^a\,
\overline{\partial}_{\mu}
\left(
\overline{\psi}_i
\overline{\gamma}^{\mu}
\mathbb{P}_R\psi_j
\right).
\end{aligned}
\end{equation}

\item 
\textbf{Fermion-antifermion-gauge-boson-ghost vertex}:
\quad
\raisebox{-0.4\height}{
  \includegraphics[
    width=0.3\linewidth
  ]{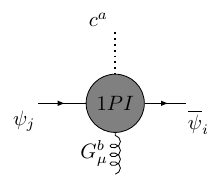}
}

\begin{equation}
\begin{aligned}
\lim_{d\rightarrow4}
\left(
\widetilde{\widehat{\Delta}}
\cdot\Gamma_{\mathrm{fin}}^{(2)}
\right)_{\psi\overline{\psi}Gc,\text{mom}}
&=
-\frac{i g_s^6}{(4\pi)^4}C_2(G)
\Bigg[
[T_{\mathcal R}^a,T_{\mathcal R}^b]
\\[0.3em]
&\quad\times
\left\{
\left(
\frac{C_2(G)}{24}
-\frac{35}{432}\sum_m C(R_m)
\right)\mathbf{1}
+\frac{C_2(\mathcal R)}{36}
\right\}
\Bigg]_{ij}
\\[0.3em]
&\quad\times
\overline{\gamma}^{\mu}\mathbb{P}_R,
\end{aligned}
\end{equation}
\begin{equation}
\begin{aligned}
\lim_{d\rightarrow4}
\left(
\widetilde{\widehat{\Delta}}
\cdot\Gamma_{\mathrm{fin}}^{(2)}
\right)_{\psi\overline{\psi}Gc}&=
\frac{g_s^6}{(4\pi)^4}C_2(G)f^{abc}
\Bigg[
T_{\mathcal R}^{c}
\\[0.3em]
&\quad\times
\left\{
\left(
\frac{C_2(G)}{24}
-\frac{35}{432}\sum_m C(R_m)
\right)\mathbf{1}
+\frac{C_2(\mathcal R)}{36}
\right\}
\Bigg]_{ij}
\\[0.5em]
&\quad\times
\int d^d x\,c^a\,
\overline{\psi}_i
\overline{\slashed{G}}^{\,b}
\mathbb{P}_R\psi_j .
\end{aligned}
\end{equation}

\item 
\textbf{BRST-source-fermion-double-ghost vertex}:
\quad
\raisebox{-0.4\height}{
  \includegraphics[
    width=0.3\linewidth
  ]{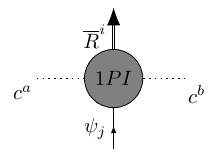}
}

\begin{equation}
\begin{aligned}
\lim_{d\rightarrow4}
\left(
\widetilde{\widehat{\Delta}}
\cdot\Gamma_{\mathrm{fin}}^{(2)}
\right)_{\overline{R}\psi cc,\text{mom}}
&=
-\frac{g_s^6}{(4\pi)^4}C_2(G)
\Bigg[
[T_{\mathcal R}^a,T_{\mathcal R}^b]
\\[0.3em]
&\quad\times
\left(
\frac{C_2(G)}{48}\mathbf{1}
-\frac{C_2(\mathcal R)}{36}
\right)
\Bigg]_{ij}\mathbb{P}_R,
\end{aligned}
\end{equation}
\begin{equation}
\begin{aligned}
\lim_{d\rightarrow4}
\left(
\widetilde{\widehat{\Delta}}
\cdot\Gamma_{\mathrm{fin}}^{(2)}
\right)_{\overline{R}\psi cc}&=
-\frac{i g_s^6}{(4\pi)^4}C_2(G)f^{abc}
\Bigg[
T_{\mathcal R}^{c}
\left(
\frac{C_2(G)}{96}\mathbf{1}
-\frac{C_2(\mathcal R)}{72}
\right)
\Bigg]_{ij}
\\[0.5em]
&\quad\times
\int d^d x\,
\overline{R}^{\,i}\mathbb{P}_R\psi_j\,c^a c^b .
\end{aligned}
\end{equation}

\item 
\textbf{BRST-source-anti-fermion-double-ghost vertex}:
\quad
\raisebox{-0.4\height}{
  \includegraphics[
    width=0.3\linewidth
  ]{
  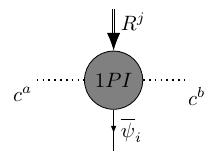
  }
}

\begin{equation}
\begin{aligned}
\lim_{d\rightarrow4}
\left(
\widetilde{\widehat{\Delta}}
\cdot\Gamma_{\mathrm{fin}}^{(2)}
\right)_{R\overline{\psi}cc,\text{mom}}
&=
-\frac{g_s^6}{(4\pi)^4}C_2(G)
\Bigg[
[T_{\mathcal R}^a,T_{\mathcal R}^b]
\\[0.3em]
&\quad\times
\left(
-\frac{C_2(G)}{48}\mathbf{1}
+\frac{C_2(\mathcal R)}{36}
\right)
\Bigg]_{ij}\mathbb{P}_L,
\end{aligned}
\end{equation}
\begin{equation}
\begin{aligned}
\lim_{d\rightarrow4}
\left(
\widetilde{\widehat{\Delta}}
\cdot\Gamma_{\mathrm{fin}}^{(2)}
\right)_{R\overline{\psi}cc}&=
-\frac{i g_s^6}{(4\pi)^4}C_2(G)f^{abc}
\Bigg[
T_{\mathcal R}^{c}
\left(
-\frac{C_2(G)}{96}\mathbf{1}
+\frac{C_2(\mathcal R)}{72}
\right)
\Bigg]_{ij}
\\[0.5em]
&\quad\times
\int d^d x\,
\overline{\psi}_i\mathbb{P}_L R^j\,c^a c^b .
\end{aligned}
\end{equation}

\item 
\textbf{Anti-ghost-ghost-ghost-vertex}:
\quad
\raisebox{-0.3\height}{
  \includegraphics[
    width=0.3\linewidth
  ]{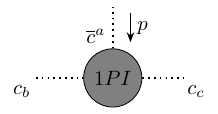}
}

\begin{equation}
\begin{aligned}
\lim_{d\rightarrow4}
\left(
\widetilde{\widehat{\Delta}}
\cdot\Gamma_{\mathrm{fin}}^{(2)}
\right)_{\overline{c}cc,\text{mom}}
&=
-\frac{g_s^5}{(4\pi)^4}\,
f^{abc}\,\frac{35}{432}C_2(G)
\sum_m C(R_m)\,\overline{p}^{\,2},
\end{aligned}
\end{equation}
\begin{equation}
\begin{aligned}
\lim_{d\rightarrow4}
\left(
\widetilde{\widehat{\Delta}}
\cdot\Gamma_{\mathrm{fin}}^{(2)}
\right)_{\overline{c}cc}&=
\frac{g_s^5}{(4\pi)^4}\,
f^{abc}\,\frac{35}{864}C_2(G)
\sum_m C(R_m)
\\[0.5em]
&\quad\times
\int d^d x\,
\overline{\partial}^{\,2}\overline{c}^{\,a}\,
c^b c^c .
\end{aligned}
\end{equation}

\item 
\textbf{Anti-ghost-ghost-ghost-gauge-boson-vertex}:
\quad
\raisebox{-0.4\height}{
  \includegraphics[
    width=0.3\linewidth
  ]{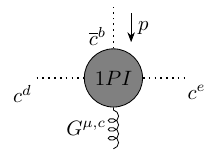}
}

\begin{equation}
\begin{aligned}
\lim_{d\rightarrow4}
\left(
\widetilde{\widehat{\Delta}}
\cdot\Gamma_{\mathrm{fin}}^{(2)}
\right)_{\overline{c}ccG,\text{mom}}
&=
\frac{i g_s^6}{(4\pi)^4}\,
f^{abc}f^{ade}\,\frac{35}{432}C_2(G)
\sum_m C(R_m)\,\overline{p}^{\,\mu},
\end{aligned}
\end{equation}
\begin{equation}
\begin{aligned}
\lim_{d\rightarrow4}
\left(
\widetilde{\widehat{\Delta}}
\cdot\Gamma_{\mathrm{fin}}^{(2)}
\right)_{\overline{c}ccG}&=
-\frac{g_s^6}{(4\pi)^4}\,
f^{abc}f^{ade}\,\frac{35}{864}C_2(G)
\sum_m C(R_m)
\\[0.5em]
&\quad\times
\int d^d x\,
\overline{\partial}^{\,\mu}\overline{c}^{\,b}\,
\overline{G}_{\mu}^{\,c}\,c^d c^e .
\end{aligned}
\end{equation}

\item 
\textbf{BRST-source-ghost-ghost-vertex}:
\quad
\raisebox{-0.3\height}{
  \includegraphics[
    width=0.3\linewidth
  ]{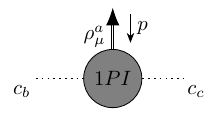}
}

\begin{equation}
\begin{aligned}
\lim_{d\rightarrow4}
\left(
\widetilde{\widehat{\Delta}}
\cdot\Gamma_{\mathrm{fin}}^{(2)}
\right)_{\rho cc,\text{mom}}
&=
-\frac{i g_s^5}{(4\pi)^4}\,
f^{abc}\,\frac{35}{432}C_2(G)
\sum_m C(R_m)\,\overline{p}^{\,\mu}.
\end{aligned}
\end{equation}
\begin{equation}
\begin{aligned}
\lim_{d\rightarrow4}
\left(
\widetilde{\widehat{\Delta}}
\cdot\Gamma_{\mathrm{fin}}^{(2)}
\right)_{\rho cc}&=
\frac{g_s^5}{(4\pi)^4}\,
f^{abc}\,\frac{35}{864}C_2(G)
\sum_m C(R_m)
\\[0.5em]
&\quad\times
\int d^d x\,
\overline{\partial}^{\,\mu}\rho_\mu^a\,
c^b c^c .
\end{aligned}
\end{equation}

\item 
\textbf{BRST-source-ghost-ghost-gauge-boson-vertex}:
\quad
\raisebox{-0.5\height}{
  \includegraphics[
    width=0.22\linewidth
  ]{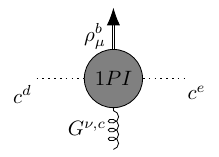}
}

\begin{equation}
\begin{aligned}
\lim_{d\rightarrow4}
\left(
\widetilde{\widehat{\Delta}}
\cdot\Gamma_{\mathrm{fin}}^{(2)}
\right)_{\rho ccG,\text{mom}}
&=
-\frac{g_s^6}{(4\pi)^4}\,
\frac{35}{432}C_2(G)\sum_m C(R_m)\,
f^{abc}f^{ade}\,\overline{g}^{\mu\nu},
\end{aligned}
\end{equation}
\begin{equation}
\begin{aligned}
\lim_{d\rightarrow4}
\left(
\widetilde{\widehat{\Delta}}
\cdot\Gamma_{\mathrm{fin}}^{(2)}
\right)_{\rho ccG}&=
-\frac{g_s^6}{(4\pi)^4}\,
\frac{35}{864}C_2(G)\sum_m C(R_m)\,
f^{abc}f^{ade}
\\[0.5em]
&\quad\times
\int d^d x\,
\rho_b^\mu\,\overline{G}_\mu^{\,c}\,c^d c^e .
\end{aligned}
\end{equation}

\end{itemize}
After determining the diagrammatic contributions to the two-loop BRST breaking, we turn to the additional terms generated by the nonlinear structure of the Slavnov--Taylor functional. At two-loop order, products of one-loop finite counter-terms contribute to the Slavnov--Taylor identity and therefore constitute an additional source of BRST breaking that must be included in the restoration condition. These contributions are given by

\begin{equation}
\begin{aligned}
\left.
\mathcal S_d\!\left(S_{\mathrm{fct}}^{(1)}\right)
\right|_{\text{2-loop}}
&=
\int d^d x\,
\left\{
\frac{\delta S_{\mathrm{fct}}^{(1)}}{\delta\psi_i}
\frac{\delta S_{\mathrm{fct}}^{(1)}}{\delta\overline R^i}
+
\frac{\delta S_{\mathrm{fct}}^{(1)}}{\delta R_i}
\frac{\delta S_{\mathrm{fct}}^{(1)}}{\delta\overline\psi^i}
\right\}
\\[0.8em]
&=
-\frac{g_s^5}{(4\pi)^4}\frac{C_2(G)}{4}
\left[C_2(\mathcal R)T_{\mathcal R}^{a}\right]_{ij}
\\[0.3em]
&\qquad\times
\int d^d x\,c^a\,
\overline{\partial}_{\mu}
\left(
\overline{\psi}_i\overline{\gamma}^{\mu}
\mathbb P_R\psi_j
\right)
\\[0.8em]
&\quad
+\frac{i g_s^6}{(4\pi)^4}
\frac{C_2(G)^2}{32}\,
f^{abc}T_{\mathcal R,ij}^{c}
\\[0.3em]
&\qquad\times
\int d^d x\,
\overline R^{\,i}\mathbb P_R\psi_j\,c^a c^b
\\[0.8em]
&\quad
-\frac{i g_s^6}{(4\pi)^4}
\frac{C_2(G)^2}{32}\,
f^{abc}T_{\mathcal R,ij}^{c}
\\[0.3em]
&\qquad\times
\int d^d x\,
\overline{\psi}_i\mathbb P_L R^j\,c^a c^b .
\end{aligned}
\end{equation}

\subsection{Finite BRST-restoring counter-terms at two loops}

We reconstruct the finite counter-terms by matching their Slavnov-Taylor variation to the breaking obtained above. The restoration identities below are understood after physical four-dimensional projection and the counter-terms retain the chosen $d$-dimensional continuation.

\begin{equation}
\begin{aligned}
&\lim_{d\rightarrow4}
\left(
\widetilde{\widehat{\Delta}}\cdot\Gamma_{\rm fin}^{(2)}
\right)_{Gc}
+
\lim_{d\rightarrow4}
\left(
\widetilde{\widehat{\Delta}}\cdot\Gamma_{\rm fin}^{(2)}
\right)_{GGc}
\\[0.5em]
&=
-s_d\Bigg\{
\frac{2g_s^4}{(4\pi)^4}
\Bigg[
-\frac{271}{4608}C_2(G)\sum_m C(R_m)
+\frac{5}{144}\sum_m C(R_m)C_2(R_m)
\Bigg]
\\[0.3em]
&\qquad\times
\int d^d x\,
\overline G_\mu^a\,
\overline\partial^2\overline G^{\mu,a}
\\[0.8em]
&\qquad
+\frac{2g_s^4}{(4\pi)^4}
\Bigg[
\frac{365}{13824}C_2(G)\sum_m C(R_m)
+\frac{23}{288}\sum_m C(R_m)C_2(R_m)
\Bigg]
\\[0.3em]
&\qquad\times
\int d^d x\,
\overline G^{\mu,a}\,
\overline\partial_\mu\overline\partial_\nu
\overline G^{\nu,a}
\Bigg\}.
\end{aligned}
\end{equation}

\begin{equation}
\begin{aligned}
&\lim_{d\rightarrow4}
\left(
\widetilde{\widehat{\Delta}}\cdot\Gamma_{\rm fin}^{(2)}
\right)_{GGGc}
\\[0.5em]
&=
-s_d\Bigg\{
\frac{g_s^6}{(4\pi)^4}
\sum_m
\Bigg[
\frac{5}{576}
\left(
\mathcal Q_{R_m}^{ad|bc}
-2\mathcal Q_{R_m}^{ab|cd}
-2\mathcal Q_{R_m}^{ac|bd}
\right)
\\[0.5em]
&\qquad
+\left(
\frac{5}{108}C_2(G)
-\frac{5}{144}C_2(R_m)
\right)\mathcal D_{R_m}^{adbc}
\\[0.5em]
&\qquad
+\left(
-\frac{235}{1728}C_2(G)
-\frac{17}{144}C_2(R_m)
\right)T_{R_m}^{abcd}
\Bigg]
\\[0.5em]
&\qquad\times
\int d^d x\,
\overline G_\mu^a\overline G^{\mu,d}
\overline G_\nu^b\overline G^{\nu,c}
\Bigg\}.
\end{aligned}
\end{equation}

For compactness, we define the following central matrices:
\begin{equation}
\begin{aligned}
\mathcal B_{\mathcal R}
&=
\left[
\frac{1}{24}C_2(G)^2
-\frac{35}{432}C_2(G)\sum_m C(R_m)
\right]\mathbf 1_{\mathcal R}
\\[0.3em]
&\quad
+\frac{1}{36}C_2(G)C_2(\mathcal R),
\\[0.8em]
\mathcal K_{\mathcal R}
&=
\left[
\frac{31}{24}C_2(G)^2
-\frac{25}{144}C_2(G)\sum_m C(R_m)
\right]\mathbf 1_{\mathcal R}
\\[0.3em]
&\quad
+\left[
\frac{41}{108}\sum_m C(R_m)
-\frac{1181}{216}C_2(G)
\right]C_2(\mathcal R)
\\[0.3em]
&\quad
+\frac{127}{36}C_2(\mathcal R)^2 .
\end{aligned}
\end{equation}

\begin{equation}
\begin{aligned}
&\lim_{d\rightarrow4}
\left(
\widetilde{\widehat{\Delta}}\cdot\Gamma_{\rm fin}^{(2)}
\right)_{\overline\psi\psi c}
+
\lim_{d\rightarrow4}
\left(
\widetilde{\widehat{\Delta}}\cdot\Gamma_{\rm fin}^{(2)}
\right)_{\overline\psi\psi Gc}
\\[0.5em]
&+
\lim_{d\rightarrow4}
\left(
\widetilde{\widehat{\Delta}}\cdot\Gamma_{\rm fin}^{(2)}
\right)_{\overline R\psi cc}
+
\lim_{d\rightarrow4}
\left(
\widetilde{\widehat{\Delta}}\cdot\Gamma_{\rm fin}^{(2)}
\right)_{\overline\psi Rcc}
\\[0.5em]
&+
\lim_{d\rightarrow4}
\left(
\widetilde{\widehat{\Delta}}\cdot\Gamma_{\rm fin}^{(2)}
\right)_{\overline ccc}
+
\lim_{d\rightarrow4}
\left(
\widetilde{\widehat{\Delta}}\cdot\Gamma_{\rm fin}^{(2)}
\right)_{\overline cccG}
\\[0.5em]
&+
\lim_{d\rightarrow4}
\left(
\widetilde{\widehat{\Delta}}\cdot\Gamma_{\rm fin}^{(2)}
\right)_{\rho cc}
+
\lim_{d\rightarrow4}
\left(
\widetilde{\widehat{\Delta}}\cdot\Gamma_{\rm fin}^{(2)}
\right)_{\rho ccG}
\\[0.5em]
&+
\left.
\mathcal S_d\!\left(S_{\rm fct}^{(1)}\right)
\right|_{\text{2-loop}}
\\[0.8em]
&=
-\mathcal S_d\Bigg\{
S_0
+\frac{g_s^4}{(4\pi)^4}
\frac{35}{432}C_2(G)\sum_m C(R_m)
\int d^d x\,\zeta^a s_d c^a
\\[0.5em]
&\qquad
-\frac{g_s^4}{(4\pi)^4}
\int d^d x\,
\Big(
\overline R^{\,i}
[\mathcal B_{\mathcal R}]_{ij}\,s_d\psi_j
+
R^j[\mathcal B_{\mathcal R}]_{ij}\,
s_d\overline\psi_i
\Big)
\\[0.5em]
&\qquad
-\frac{i g_s^4}{(4\pi)^4}
\int d^d x\,
\overline\psi_i
[\mathcal K_{\mathcal R}]_{ij}\,
\overline{\slashed{\partial}}\mathbb P_R\psi_j
\Bigg\}\Bigg|_{\text{2-loop}} .
\end{aligned}
\end{equation}

In addition to the source-variation identity used above,
we employ the following identity. The contribution quadratic in
$\int d^dx\,\zeta^a s_dc^a$ vanishes by the Jacobi identity,
so no additional loop-order projection is needed in this
unweighted source identity:

\begin{equation}
\label{eqn:zeta-fct-var}
\begin{aligned}
&\mathcal S_d
\Bigg\{
S_0+\int d^d x\,\zeta^a s_d c^a
\Bigg\}\Bigg|_{\mathrm{4D}}
\\[0.5em]
&=
-\frac{g_s}{2}f^{abc}
\int d^d x\,
\overline\partial^2\overline c^{\,a}c^b c^c
\\[0.5em]
&\quad
+\frac{g_s^2}{2}f^{abc}f^{ade}
\int d^d x\,
\overline\partial^\mu\overline c^{\,b}
G_\mu^c c^d c^e
\\[0.5em]
&\quad
-\frac{g_s}{2}f^{abc}
\int d^d x\,
\overline\partial_\mu\rho_a^\mu c^b c^c
\\[0.5em]
&\quad
+\frac{g_s^2}{2}f^{abc}f^{ade}
\int d^d x\,
\rho_b^\mu G_\mu^c c^d c^e
\\[0.5em]
&\quad
-\frac{i g_s^2}{2}f^{abc}T_{\mathcal R,ij}^c
\int d^d x\,
\overline R^{\,i}c^a c^b\mathbb P_R\psi_j
\\[0.5em]
&\quad
+\frac{i g_s^2}{2}f^{abc}T_{\mathcal R,ij}^c
\int d^d x\,
\overline\psi_i\mathbb P_L c^a c^b R_j .
\end{aligned}
\end{equation}
The appearance of the BRST sources $R$ and $\overline{R}$ in the variations of both Eqs.~\eqref{eqn:R-fct-var} and \eqref{eqn:zeta-fct-var} implies that these contributions cannot be treated independently and must instead be combined when solving the restoration condition. A similar interplay occurs for the BRST variation of the fermionic kinetic term, part of which is compensated by the variation of the contribution in Eq.~\eqref{eqn:R-fct-var}. The solution to the BRST-restoration condition therefore naturally decomposes into

\begin{equation}
\begin{aligned}
S_{\rm fct}^{(2)}
&=
\frac{2g_s^4}{(4\pi)^4}
\Bigg[
-\frac{271}{4608}C_2(G)\sum_m C(R_m)
+\frac{5}{144}\sum_m C(R_m)C_2(R_m)
\Bigg]
\\[0.3em]
&\qquad\times
\int d^d x\,
\overline G_\mu^a\overline\partial^2\overline G^{\mu,a}
\\[0.8em]
&\quad
+\frac{2g_s^4}{(4\pi)^4}
\Bigg[
\frac{365}{13824}C_2(G)\sum_m C(R_m)
+\frac{23}{288}\sum_m C(R_m)C_2(R_m)
\Bigg]
\\[0.3em]
&\qquad\times
\int d^d x\,
\overline G^{\mu,a}
\overline\partial_\mu\overline\partial_\nu
\overline G^{\nu,a}
\\[0.8em]
&\quad
+\frac{g_s^6}{(4\pi)^4}\sum_m
\Bigg[
\frac{5}{576}
\left(
\mathcal Q_{R_m}^{ad|bc}
-2\mathcal Q_{R_m}^{ab|cd}
-2\mathcal Q_{R_m}^{ac|bd}
\right)
\\[0.5em]
&\qquad
+\left(
\frac{5}{108}C_2(G)
-\frac{5}{144}C_2(R_m)
\right)\mathcal D_{R_m}^{adbc}
\\[0.5em]
&\qquad
+\left(
-\frac{235}{1728}C_2(G)
-\frac{17}{144}C_2(R_m)
\right)T_{R_m}^{abcd}
\Bigg]
\\[0.3em]
&\qquad\times
\int d^d x\,
\overline G_\mu^a\overline G^{\mu,d}
\overline G_\nu^b\overline G^{\nu,c}
\\[0.8em]
&\quad
+\frac{g_s^4}{(4\pi)^4}
\frac{35}{432}C_2(G)\sum_m C(R_m)
\int d^d x\,\zeta^a s_d c^a
\\[0.8em]
&\quad
-\frac{g_s^4}{(4\pi)^4}
\int d^d x\,
\Big(
\overline R^{\,i}
[\mathcal B_{\mathcal R}]_{ij}\,s_d\psi_j
+
R^j[\mathcal B_{\mathcal R}]_{ij}\,
s_d\overline\psi_i
\Big)
\\[0.8em]
&\quad
-\frac{i g_s^4}{(4\pi)^4}
\int d^d x\,
\overline\psi_i
[\mathcal K_{\mathcal R}]_{ij}\,
\overline{\slashed{\partial}}\mathbb P_R\psi_j .
\end{aligned}
\end{equation}
where $\mathcal D_{R_m}$ and $\mathcal Q_{R_m}$ are the trace
tensors defined previously, evaluated in each irreducible block
$R_m$ of $\mathcal R=\bigoplus_m R_m$.
In the $SU(2)$ limit, with $C_2(G)=2$ and
$S_2(R_m)=C(R_m)$ in the notation of
Ref.~\cite{Kuhler:2025znv}, the fermion and external-source sectors
agree directly after matching conventions.
In the physical four-dimensional limit, the gauge-field sectors
differ by the gauge-invariant finite counter-term
\begin{equation}
\begin{aligned}
S_{\rm fct,KS}^{(2)}-S_{\rm fct}^{(2)}
&=
-\frac{g_s^4}{(4\pi)^4}
\Bigg[
\frac{365}{6912}\sum_m C(R_m)
\\
&\qquad
+\frac{23}{288}\sum_m C(R_m)C_2(R_m)
\Bigg]
\int d^4x\,
\overline G_{\mu\nu}^{a}\overline G^{a,\mu\nu}.
\end{aligned}
\end{equation}
This single finite shift accounts for the differences
in the quadratic, cubic, and quartic gauge-field terms.
The comparison holds for arbitrary irreducible blocks
of the reducible $SU(2)$ representation.

\subsection{Pole consistency check at two loops}
We have performed several algebraic checks of the divergent
two-loop BRST-breaking contributions. In the $Gc$,
$\psi\overline\psi c$, $\overline R\psi cc$, and
$R\overline\psi cc$ channels, we have verified that the poles
obtained from the insertion diagrams are exactly
compensated by the Slavnov--Taylor variation of the
two-loop singular counterterms, together with the
divergent part of the contribution quadratic in the
one-loop counterterms. The latter includes products
of singular and finite one-loop counterterms.
These checks retain both simple and double poles,
including the evanescent tensor structures.

We have also verified the anti-ghost identities relating the
divergent contributions in the $\overline ccc$ and
$\rho cc$ channels, and in the $\overline cGcc$ and
$\rho Gcc$ channels, retaining both barred and hatted
components. Furthermore, the divergent $\zeta ccc$
contribution vanishes by colour antisymmetry and
dummy-index relabelling, without invoking anomaly
cancellation.

To illustrate the pole-restoration check explicitly,
we present the gauge-boson--ghost channel in detail
below, displaying the divergent breaking and the
counter-term contributions that cancel it.

\textbf{Gauge-boson-ghost function}:
\quad
\raisebox{-0.4\height}{
  \includegraphics[
    width=0.18\linewidth
  ]{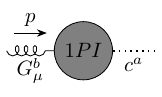}
}

\begin{equation}
\begin{aligned}
\left(
\widetilde{\widehat{\Delta}}
\cdot\Gamma_{\rm div}^{(2)}
\right)_{Gc,\text{mom}}
&=
-\frac{2i g_s^4}{(4\pi)^4}\delta^{ab}
\Bigg\{
\sum_m C(R_m)C_2(R_m)
\left(
\frac{1}{12\epsilon^2}
-\frac{17}{144\epsilon}
\right)
\widehat p^{\,2}\overline p^\mu
\\[0.5em]
&\quad
+C_2(G)\sum_m C(R_m)
\Bigg[
\left(
\frac{11}{144\epsilon^2}
-\frac{17}{1728\epsilon}
\right)
\overline p^{\,2}\widehat p^\mu
\\[0.5em]
&\qquad
+\left(
-\frac{1}{6\epsilon^2}
-\frac{1}{4\epsilon}
\right)
\widehat p^{\,2}\overline p^\mu
-\frac{11}{72\epsilon}
\overline p^{\,2}\overline p^\mu
\\[0.5em]
&\qquad
+\left(
-\frac{1}{72\epsilon^2}
-\frac{5}{864\epsilon}
\right)
\widehat p^{\,2}\widehat p^\mu
\Bigg]
\Bigg\},
\end{aligned}
\end{equation}
\begin{equation}
\begin{aligned}
\left(
\widetilde{\widehat{\Delta}}
\cdot\Gamma_{\rm div}^{(2)}
\right)_{Gc}
&=
-\frac{2g_s^4}{(4\pi)^4}
\int d^d x\,\delta^{ab}c^a
\Bigg\{
\sum_m C(R_m)C_2(R_m)
\\[0.3em]
&\qquad\times
\left(
\frac{1}{12\epsilon^2}
-\frac{17}{144\epsilon}
\right)
\widehat\partial^{\,2}\overline\partial^\mu
\\[0.5em]
&\quad
+C_2(G)\sum_m C(R_m)
\Bigg[
\left(
\frac{11}{144\epsilon^2}
-\frac{17}{1728\epsilon}
\right)
\overline\partial^{\,2}\widehat\partial^\mu
\\[0.5em]
&\qquad
+\left(
-\frac{1}{6\epsilon^2}
-\frac{1}{4\epsilon}
\right)
\widehat\partial^{\,2}\overline\partial^\mu
-\frac{11}{72\epsilon}
\overline\partial^{\,2}\overline\partial^\mu
\\[0.5em]
&\qquad
+\left(
-\frac{1}{72\epsilon^2}
-\frac{5}{864\epsilon}
\right)
\widehat\partial^{\,2}\widehat\partial^\mu
\Bigg]
\Bigg\}G_\mu^b .
\end{aligned}
\end{equation}

To determine the BRST breaking associated with this contribution at two-loop order, we evaluate the relevant terms directly in position space. The calculation naturally separates into two contributions:

\begin{enumerate}
\item
\begin{equation}
\begin{aligned}
&\int d^d x\,
\left.
\left(
\frac{\delta S_0}{\delta\rho_a^\mu}
\frac{\delta S_{\rm sct}^{(2)}}{\delta G_\mu^a}
+
\frac{\delta S_0}{\delta G_\mu^a}
\frac{\delta S_{\rm sct}^{(2)}}{\delta\rho_a^\mu}
\right)
\right|_{Gc}
\\[0.8em]
&=
\frac{2g_s^4}{(4\pi)^4}
\int d^d x\,\delta^{ab}c^a
\Bigg\{
\sum_m C(R_m)C_2(R_m)
\\[0.3em]
&\qquad\times
\left(
\frac{1}{12\epsilon^2}
-\frac{17}{144\epsilon}
\right)
\widehat\partial^{\,2}\overline\partial^\mu
\\[0.5em]
&\qquad
+C_2(G)\sum_m C(R_m)
\Bigg[
\left(
\frac{11}{144\epsilon^2}
-\frac{17}{1728\epsilon}
\right)
\overline\partial^{\,2}\widehat\partial^\mu
\\[0.5em]
&\qquad\qquad
+\left(
-\frac{1}{4\epsilon^2}
-\frac{1}{4\epsilon}
\right)
\widehat\partial^{\,2}\overline\partial^\mu
-\frac{17}{72\epsilon}
\overline\partial^{\,2}\overline\partial^\mu
\\[0.5em]
&\qquad\qquad
+\left(
-\frac{1}{72\epsilon^2}
-\frac{5}{864\epsilon}
\right)
\widehat\partial^{\,2}\widehat\partial^\mu
\Bigg]
\Bigg\}G_\mu^b .
\end{aligned}
\end{equation}

\item
\begin{equation}
\begin{aligned}
&\int d^d x\,
\left.
\left(
\frac{\delta S_{\rm ct}^{(1)}}{\delta\rho_a^\mu}
\frac{\delta S_{\rm ct}^{(1)}}{\delta G_\mu^a}
\right)
\right|_{Gc}^{\rm div}
\\[0.5em]
&=
\frac{2g_s^4}{(4\pi)^4}
C_2(G)\sum_m C(R_m)
\int d^d x\,\delta^{ab}c^a
\\[0.3em]
&\qquad\times
\left[
\frac{1}{12\epsilon^2}
\widehat\partial^{\,2}\overline\partial^\mu
+
\frac{1}{12\epsilon}
\overline\partial^{\,2}\overline\partial^\mu
\right]G_\mu^b .
\end{aligned}
\end{equation}
In the expression above, $S_{\rm ct}^{(1)}$ denotes
$S_{\rm sct}^{(1)}+S_{\rm fct}^{(1)}$ and only the overall
singular contribution is retained.
\end{enumerate}

Indeed, these two contributions combine to yield
\begin{equation}
\left.
\mathcal S_d
\left(
S_0+S_{\rm ct}^{(1)}+S_{\rm sct}^{(2)}
\right)
\right|_{Gc}^{\rm 2-loop,div}
=
-\left(
\widetilde{\widehat{\Delta}}
\cdot\Gamma_{\rm div}^{(2)}
\right)_{Gc}.
\end{equation}

The pole cancellations in the tested channels, together
with the antighost identities, provide nontrivial checks
of the two-loop counterterm structure, including its
evanescent components and the contributions quadratic
in one-loop counterterms. Combined with the finite
restoration conditions derived above, these checks
support the consistency of our two-loop renormalization
and BRST-restoration procedure in Feynman gauge.
\section{Conclusions}

We have carried out the renormalization and BRST-symmetry restoration of massless chiral \(SU(N)\) gauge theories without scalar fields through two-loop order in the Breitenlohner--Maison--'t Hooft–-Veltman scheme with fermions  in a generic, possibly reducible, representation, subject to cancellation of the perturbative gauge anomaly. We have adopted a definite \(d\)-dimensional continuation in which the chiral fermion-–gauge interaction is restricted to the physical four-dimensional Lorentz subspace. The one-loop calculation has been performed for a general covariant-gauge parameter \(\xi\), while the two-loop analysis has been restricted to Feynman gauge.

The breaking of BRST invariance induced by the BMHV continuation has been treated using the Quantum Action Principle. The resulting local breaking terms were obtained from Green functions with insertions of purely evanescent lower-level BRST-breaking operators. Ultraviolet divergences were isolated with an \(R\)-operation supplemented by an auxiliary infrared mass, while the dependence on evanescent Lorentz structures was retained separately in order to identify finite contributions generated by their interplay with ultraviolet poles.

 One loop results agree with~\cite{BeluscaMaito_2020} and partially with~\cite{Martin:1999cc}.  The reduction to \(SU(2)\) agrees with the results of~\cite{Kuhler:2025znv}, up to finite BRST-invariant terms. The UV poles have been computed directly and through local insertion operators, providing an additional nontrivial consistency check of the interplay between genuine two-loop contributions, one-loop counter-term insertions in one-loop diagrams and the nonlinear terms of the Slavnov–Taylor functional.

It is evident that the symmetry-restoring action is not unique. Finite BRST-invariant local operators can be added without affecting the Slavnov–Taylor identity, and evanescent modifications of the \(d\)-dimensional action or of lower-order counterterms can modify the representative required at higher orders. The two-loop finite counterterms presented here should therefore be understood as defining a particular BMHV renormalization prescription, specified by our choice for the  \(d\)-dimensional continuation of the tree-level action, our UV subtraction procedure and choice of one-loop finite counter-terms used in two-loop diagrams. Physical quantities are insensitive to this freedom once the corresponding scheme is implemented consistently.

The present calculation extends the explicit multi-loop implementation of the BMHV scheme from the previously studied Abelian and \(SU(2)\) examples to general chiral \(SU(N)\) matter representations. It also provides a basis for further applications in more realistic theories. Natural extensions include a calculation at general covariant-gauge parameter, the inclusion of scalar and Yukawa interactions, and ultimately applications to the Standard Model and its effective field theories. The automation of the insertion, subtraction and group-theory reduction procedures developed here should be particularly useful for such higher-loop calculations.

\section*{Acknowledgements}
This research was funded by the Swiss National Science Foundation (SNSF) [Grant number: 10003654 - Effective Field Theories beyond NLO].
\clearpage
\printbibliography[heading=bibintoc]
\appendix 

\section{Group-theory conventions}
We collect here the group-theory conventions used throughout this
work. The generators of $SU(N)$ in an irreducible representation $r$
are denoted by $T_r^a$ and satisfy
\begin{equation}
[T_r^a,T_r^b]=i f^{abc}T_r^c,
\end{equation}
where $f^{abc}$ are the structure constants of the gauge group.
We use Hermitian generators and the normalization $C(F)=1/2$
for the fundamental representation.

The Dynkin index $C(r)$ and the quadratic Casimir $C_2(r)$
are defined by
\begin{equation}
\operatorname{Tr}_r(T_r^aT_r^b)=C(r)\delta^{ab},
\qquad
\sum_a T_r^aT_r^a=C_2(r)\mathbf 1_r.
\end{equation}
They satisfy
\begin{equation}
C_2(r)d(r)=C(r)d(G),
\end{equation}
where $d(r)$ is the dimension of the representation and
$d(G)=N^2-1$. Furthermore,
\begin{equation}
\operatorname{Tr}_r(\mathbf 1_r)=d(r),
\qquad
\operatorname{Tr}_r(T_r^a)=0.
\end{equation}

For the adjoint representation, denoted by $A$, we have
\begin{equation}
(T_A^a)_{bc}=-i f^{abc},
\qquad
f^{acd}f^{bcd}=C_2(G)\delta^{ab},
\qquad
C(A)=C_2(A)=C_2(G)=N.
\end{equation}
Repeated adjoint indices are summed. We retain $C_2(G)$ and
$d(G)$ symbolically in the results rather than substituting
their explicit $SU(N)$ values.

For the reducible fermion representation introduced above,
with repeated irreducible copies counted as separate blocks,
we have
\begin{equation}
\begin{aligned}
d(\mathcal R)&=\sum_m d(R_m),
\\
C(\mathcal R)&\equiv\sum_m C(R_m),
\\
\operatorname{Tr}_{\mathcal R}
\left(T_{\mathcal R}^aT_{\mathcal R}^b\right)
&=C(\mathcal R)\delta^{ab}.
\end{aligned}
\end{equation}

In contrast to the Dynkin index, the quadratic Casimir of the
full multiplet is generally a matrix:
\begin{equation}
\begin{aligned}
C_2(\mathcal R)
&\equiv
\sum_a T_{\mathcal R}^aT_{\mathcal R}^a
\\
&=
\operatorname{diag}
\left(
C_2(R_1)\mathbf 1_{R_1},\ldots,
C_2(R_M)\mathbf 1_{R_M}
\right),
\\
[C_2(\mathcal R),T_{\mathcal R}^a]&=0.
\end{aligned}
\end{equation}
It must therefore be retained as an operator on open fermion
lines, rather than replaced by a common scalar. The dimension
identity generalizes to
\begin{equation}
\operatorname{Tr}_{\mathcal R}C_2(\mathcal R)
=
\sum_m C_2(R_m)d(R_m)
=
d(G)\sum_m C(R_m).
\end{equation}

The colour reduction uses consequences of the commutation
relations, trace cyclicity,
\begin{equation}
\operatorname{Tr}_X(T_X^{a_1}\cdots T_X^{a_n})
=
\operatorname{Tr}_X(T_X^{a_2}\cdots T_X^{a_n}T_X^{a_1}),
\end{equation}
and contracted-generator identities. In the reducible
representation, these include
\begin{equation}
\begin{aligned}
\sum_a T_{\mathcal R}^aT_{\mathcal R}^bT_{\mathcal R}^a
&=
\left(
C_2(\mathcal R)-\frac{C_2(G)}{2}\mathbf 1_{\mathcal R}
\right)T_{\mathcal R}^b,
\\[0.5em]
f^{abc}T_{\mathcal R}^aT_{\mathcal R}^b
&=
\frac{i}{2}C_2(G)T_{\mathcal R}^c.
\end{aligned}
\end{equation}
The crossed contraction used at two loops is
\begin{equation}
\begin{aligned}
\sum_{a,b}
T_{\mathcal R}^aT_{\mathcal R}^bT_{\mathcal R}^c
T_{\mathcal R}^aT_{\mathcal R}^b
&=
\Bigg[
C_2(\mathcal R)^2
-\frac32 C_2(G)C_2(\mathcal R)
\\
&\qquad
+\frac12 C_2(G)^2\mathbf 1_{\mathcal R}
\Bigg]T_{\mathcal R}^c.
\end{aligned}
\end{equation}

For products of structure constants, the relevant identities
include the quadratic adjoint contraction above and
\begin{equation}
\begin{aligned}
f^{abe}f^{cde}
-f^{ace}f^{bde}
+f^{ade}f^{bce}
&=0,
\\[0.5em]
f^{axy}f^{bzy}f^{czx}
&=-\frac{C_2(G)}{2}f^{abc}.
\end{aligned}
\end{equation}

Closed fermion chains with Casimir insertions give weighted
traces. For $k\geq0$, define
\begin{equation}
\begin{aligned}
\mathcal T_k^{a_1\cdots a_n}
&\equiv
\operatorname{Tr}_{\mathcal R}
\left(
C_2(\mathcal R)^k
T_{\mathcal R}^{a_1}\cdots T_{\mathcal R}^{a_n}
\right)
\\
&=
\sum_m C_2(R_m)^k
\operatorname{Tr}_{R_m}
\left(
T_{R_m}^{a_1}\cdots T_{R_m}^{a_n}
\right).
\end{aligned}
\end{equation}
In particular,
\begin{equation}
\begin{aligned}
\operatorname{Tr}_{\mathcal R}
\left(C_2(\mathcal R)^kT_{\mathcal R}^a\right)
&=0,
\\[0.5em]
\operatorname{Tr}_{\mathcal R}
\left(
C_2(\mathcal R)^kT_{\mathcal R}^aT_{\mathcal R}^b
\right)
&=
\left[\sum_m C_2(R_m)^kC(R_m)\right]\delta^{ab}.
\end{aligned}
\end{equation}
For positive integer $k$, an empty weighted trace satisfies
\begin{equation}
\operatorname{Tr}_{\mathcal R}C_2(\mathcal R)^k
=
d(G)\sum_m C_2(R_m)^{k-1}C(R_m).
\end{equation}
Thus the sums $\sum_m C(R_m)$ and
$\sum_m C_2(R_m)C(R_m)$ are distinct invariants.
Separate closed fermion loops carry independent block sums.

We impose perturbative gauge-anomaly cancellation on the
total fermion representation:
\begin{equation}
\begin{aligned}
\operatorname{Tr}_{\mathcal R}
\left(
T_{\mathcal R}^a
\left\{T_{\mathcal R}^b,T_{\mathcal R}^c\right\}
\right)
&=
\sum_m
\operatorname{Tr}_{R_m}
\left(
T_{R_m}^a
\left\{T_{R_m}^b,T_{R_m}^c\right\}
\right)
=0.
\end{aligned}
\end{equation}
This is an additional restriction on the matter content,
not an identity satisfied by every representation.
Individual blocks need not be anomaly-free.
\par
For $X=\mathcal R,R_m,A$, the cubic trace decomposes as
\begin{equation}
\begin{aligned}
\operatorname{Tr}_X(T_X^aT_X^bT_X^c)
&=
d_X^{abc}+\frac{i}{2}C(X)f^{abc},
\\[0.5em]
d_X^{abc}
&\equiv
\operatorname{Tr}_X
\left(T_X^{(a}T_X^bT_X^{c)}\right).
\end{aligned}
\end{equation}
Parentheses denote normalized symmetrization over all enclosed
indices. For $d_X^{abc}$, this means symmetrization over
$a,b,c$ with weight $1/3!$. Trace cyclicity reduces the six
terms to two distinct cyclic orderings, each with coefficient
$1/2$.
Explicitly,
\begin{equation}
\begin{aligned}
d_X^{abc}
&=
\frac{1}{3!}\sum_{\pi\in S_3}
\operatorname{Tr}_X
\left(
T_X^{a_{\pi(1)}}T_X^{a_{\pi(2)}}T_X^{a_{\pi(3)}}
\right),
\qquad (a_1,a_2,a_3)=(a,b,c),
\\[0.5em]
&=
\frac12\operatorname{Tr}_X
\left(
T_X^a\left\{T_X^b,T_X^c\right\}
\right).
\end{aligned}
\end{equation}
The anomaly-cancellation condition is therefore equivalent to
\begin{equation}
d_{\mathcal R}^{abc}
=
\sum_m d_{R_m}^{abc}
=0,
\end{equation}
while $d_A^{abc}=0$ identically. Accordingly,
\begin{equation}
\operatorname{Tr}_{\mathcal R}
\left(T_{\mathcal R}^aT_{\mathcal R}^bT_{\mathcal R}^c\right)
=
\frac{i}{2}\sum_m C(R_m)f^{abc}.
\end{equation}
This condition does not in general imply
\begin{equation}
\sum_m C_2(R_m)^k d_{R_m}^{abc}=0,
\qquad k>0.
\end{equation}
Casimir-weighted symmetric cubic traces are therefore retained.

We denote ordered quartic traces by
\begin{equation}
T_X^{abcd}
=
\operatorname{Tr}_X(T_X^aT_X^bT_X^cT_X^d),
\qquad X=\mathcal R,R_m,A,
\end{equation}
with
\begin{equation}
T_{\mathcal R}^{abcd}=\sum_m T_{R_m}^{abcd}.
\end{equation}
Closed adjoint chains can be written as
\begin{equation}
f^{aef}f^{bfg}f^{cgh}f^{dhe}=T_A^{abcd}.
\end{equation}
Different trace orderings are related by adjacent
commutations, for example,
\begin{equation}
T_X^{abcd}-T_X^{bacd}
=
i f^{abe}\operatorname{Tr}_X(T_X^eT_X^cT_X^d).
\end{equation}

The symmetrized quartic trace is
\begin{equation}
\begin{aligned}
d_X^{abcd}
&\equiv
\operatorname{Tr}_X
\left(T_X^{(a}T_X^bT_X^cT_X^{d)}\right)
\\[0.5em]
&=
\frac16
\Big(
T_X^{abcd}+T_X^{abdc}
+T_X^{acbd}+T_X^{acdb}
+T_X^{adbc}+T_X^{adcb}
\Big),
\end{aligned}
\end{equation}
where symmetrization is over all four indices with weight
$1/4!$ and cyclicity gives the six-ordering expression above.
In particular,
\begin{equation}
d_{\mathcal R}^{abcd}=\sum_m d_{R_m}^{abcd}.
\end{equation}
The same cyclicity and commutator identities apply to
Casimir-weighted traces because $C_2(\mathcal R)$ commutes
with every generator.

\end{document}